\documentclass[%
 reprint,
superscriptaddress,
 amsmath,amssymb,
 aps,
prb,
]{revtex4-2}

\usepackage{graphicx}
\usepackage{dcolumn}
\usepackage{bm}

\usepackage[utf8]{inputenc}
\usepackage[T1]{fontenc}
\usepackage{mathptmx}
\usepackage{etoolbox}
\usepackage[version=4]{mhchem}
\usepackage{siunitx}
\usepackage{amsmath}
\usepackage{lipsum}
\usepackage{array}
\usepackage{placeins}
\usepackage[normalem]{ulem} 
\usepackage{xcolor}
\usepackage[ colorlinks,
    linkcolor={blue},
    citecolor={blue},
    urlcolor={blue}]{hyperref}
\usepackage{braket}

\newcommand{\evgw}{ev$GW_0$ }
\newcommand{\evgwnospace}{ev$GW_0$}
\newcommand{\mathd}{\mathrm{d}}

\newcommand{\tmop}[1]{\ensuremath{\operatorname{#1}}}
\newcommand\eqt{\hspace{0.17em}{=}\hspace{0.17em}}
\newcommand\rightarrowtext{\hspace{0.17em}{\rightarrow}\hspace{0.17em}}
\newcommand\intext{\hspace{0.17em}{\in}\hspace{0.17em}}

\newcommand\letext{\hspace{0.17em}{\le}\hspace{0.17em}}
\newcommand\apt{\hspace{0.17em}{\approx}\hspace{0.17em}}

\newcommand\pt{\hspace{0.17em}{+}\hspace{0.17em}}
\newcommand\mt{\hspace{0.17em}{-}\hspace{0.17em}}

\newcommand{\gwoh}{\textit{GW}100~}
\newcommand{\br}{\mathbf{r}}
\newcommand{\sgn}{\text{sgn}}

\definecolor{pine}{RGB}{40, 105, 131}

\newcommand{\reviewnew}[1]{#1}
\newcommand{\reviewout}[1]{}

\newcommand{\fbnew}[1]{{#1}}
\newcommand{\fbout}[1]{}

\begin{document}

\preprint{APS/123-QED}

\title{Solving multi-pole challenges in the \gwoh benchmark  enables precise low-scaling \textit{GW} calculations}

\author{Mia Schambeck}
\affiliation{Institute  of  Theoretical  Physics and Regensburg Center for Ultrafast Nanoscopy (RUN),  University  of  Regensburg,  93053  Regensburg,  Germany}
\author{Dorothea Golze}
\affiliation{Faculty for Chemistry and Food Chemistry, Technische Universit\"at Dresden, 01062 Dresden, Germany}
\author{Jan Wilhelm}
 \email{jan.wilhelm@physik.uni-regensburg.de}

\affiliation{Institute  of  Theoretical  Physics and Regensburg Center for Ultrafast Nanoscopy (RUN),  University  of  Regensburg,  93053  Regensburg,  Germany}

\date{\today}

\begin{abstract}
The $GW$ approximation is a widely used method for computing electron addition and removal energies of molecules and solids.
The computational effort of conventional $GW$ algorithms increases as $O(N^4)$ with the system size~$N$, hindering the application of $GW$ to large and complex systems. 
Low-scaling $GW$ algorithms are currently very actively developed. Benchmark studies at the single-shot $G_0W_0$ level indicate excellent numerical precision for frontier quasiparticle energies, with mean absolute deviations $<10$~meV \reviewnew{between low-scaling and} \reviewout{with respect to} standard implementations for the widely used \gwoh test set. 
A notable challenge for low-scaling $GW$ algorithms remains in achieving high precision for five molecules within the \gwoh test set, namely \ce{O3}, \ce{BeO}, \ce{MgO}, \ce{BN}, and \ce{CuCN}, for which the deviations are in the range of several hundred meV \reviewnew{at the $G_0W_0$ level}.
This is due to a spurious transfer of spectral weight from the quasiparticle to the satellite spectrum 
 in $G_0W_0$ calculations, resulting in multi-pole features in the self-energy and spectral function, which low-scaling algorithms fail to describe.
We show in this work that including eigenvalue self-consistency in the Green's function (\evgwnospace) achieves a proper separation between satellite and quasiparticle peak, leading to a single  solution of the quasiparticle equation with spectral weight close to one.
\evgw quasiparticles energies from low-scaling $GW$ closely align with reference calculations; the mean absolute error is only 12~meV for the five molecules. 
\reviewnew{We thus demonstrate that  low-scaling $GW$ with self-consistency in $G$  is well-suited for computing frontier quasiparticle energies.}

\end{abstract}

\maketitle

\section{Introduction}

\textit{GW} calculations~\cite{hedin1965new,Golze2019,Reining2018} have become a standard  tool for calculating electron addition and removal energies of molecules~\cite{Blase2011,gw100,Knight2016}, two-dimensional materials~\cite{Molina2013,Xia2020,Gjerding2021recent,Rasmussen2021,Mitterreiter2021,Lin2021,Guandalini2023efficient,graml2023lowscaling,Camarasa2023,Camarasa2024,RodriguesPela2024,Krumland2024,Krumland2024a} and solids~\cite{bruneval2008accurate,Filip2014,Cho2019,Ren/etal:2021,Biega2023,Biffi2023,Leppert2024}.  
 Recent advancements of the $GW$ method span a broad spectrum,  including the application to deep core excitations~\cite{Golze2018,Golze2020,keller2020relativistic,zhu2021all,mejia2021,Mejia2022basis,li2022benchmark,panades2023,Galleni2024}, relativistic \textit{GW} schemes~\cite{Kuehn2015one,holzer2019ionized,Yeh2022,foerster2023twocomponent,kehry2023,Gaurav2024,abraham2024}, exploring excited-state potential energy surfaces from \textit{GW}+Bethe-Salpeter~\cite{Caylak2021,Berger2021,knysh2023excited,Knysh2024},   electron dynamics from Green's functions~\cite{attaccalite2011realtime,jiang2021realtime,chan2021giant,Schluenzen2020achieving,Joost2020,Perfetto2022realtime,Tuovinen2020comparing,tuovinen2021electronic,tuovinen2023,Pavlyukh2023,Bonitz2024,Reeves2024}, and applying the \textit{GW} methodology in magnetic fields~\cite{holzer2019GW,holzer2021gw,franzke2022nmr,holzer2023practical}. 
There has also been a concerted effort towards benchmarking the accuracy of the \textit{GW} method~\cite{bruneval2021gw,Orlando2023,Marek2024,Ammar2024} and the numerical precision of $GW$ implementations~\cite{gw100,maggio2017gw100,setten2017,gao2019real,govoni2018,foerster2021gw100,rangel2020reproducibility}.
An increasing number of machine learning models  have emerged for predicting quasiparticle (QP) energies without performing a costly $GW$ calculation~\cite{stuke2020atomic,Westermayr2021,Golze2022,Fediai2023,Mondal2023,Zauchner2023,Venturella2024}.
Concerning method development beyond the $GW$ approximation,  vertex corrections~\cite{DelSole1994,Shirley1995,Bruneval2005,Shishkin2007,Grueneis2014,Maggio2017Vertex,Lewis2019,Vlcek2019,Tal2021,wangy2021,Foerster2022ExploringG3W2,Lorin2023,bruneval2024fully,Wen2024} and connections with coupled-cluster methods~\cite{lange2018,toelle2023,tölle2023abg0w0} have been explored.
Promising developments for the study of electronic excitations in the solid state have been also reported using coupled-cluster methods~\cite{McClain2017,wang2020,wang2021,Laughon2022,gallo2021,masios2023,Robinson2024} and low-scaling Bethe-Salpeter approaches~\cite{Fuchs2008,Blase2020,Merkel2024}.
Despite that $GW$ method development is blooming, several algorithmic bottlenecks render \textit{GW} calculations challenging, particularly in dealing with complex or disordered systems with large simulation cells. 
In  conventional \textit{GW} implementations,  the computational cost increases  as $O(N^4)$ with the system size~$N$, restricting conventional \textit{GW} calculations often to systems with a few hundred atoms~\cite{Golze2019,stuke2020atomic}. 
Various strategies have been devised to tackle this limitation, ranging from massively parallel implementations~\cite{govoni2015large,Kim2019scalable,Sangalli2019many,del2019large,delben2020accelerating,yu2020gpu,yeh2022fully}, to physically motivated approximations like embedding~\cite{duchemin2016combining,Li2016,Li2018,toelle2021subsystem,amblard2023manybody,amblard2024} and low-scaling techniques~\cite{rojas_1995,rieger1999gw,foerster2011,stochastic_gw,liu2016,vlcek2017stochastic,Vlcek2018,Gao2018,wilhelm2018,duchemin2019,foerster2020,Kim2020,Kutepov2020,Gao2022,wilhelm2021,Yeh2024,Shi2024}. 
In this work, we focus on deterministic, low-scaling $GW$ algorithms based on the \textit{GW} space-time method by Rojas, Godby and Needs published in 1995~\cite{rojas_1995}.
The $GW$ space-time method relies on space-local representations and imaginary time-frequency transforms and achieves cubic scaling of the computational cost in the system size, ${O}({N^3})$, instead of quartic scaling ${O}({N^4})$ of conventional $GW$ algorithms.
Many different techniques have been used to increase the computational efficiency of the $GW$ space-time method, including pair-atomic resolution of the identity~\cite{foerster2020,forster2021low,foerster2022quasiparticle,spadetto2023toward,golze2017LRI}, separable density fitting~\cite{Lu2015,Lu2017,duchemin2019,duchemin2021,Delesma2024,Yeh2024,Gao2024}, and global resolution of the identiy with a local metric~\cite{Vahtras1993,Jung2005,wilhelm2018,wilhelm2021,graml2023lowscaling}.
All of the space-time $GW$ algorithms include three Fourier transforms  between imaginary time and imaginary frequency, and vice versa, which are  performed numerically. 
This poses a significant challenge in terms of numerical precision, a challenge that has been addressed by the development of tailored imaginary-time and imaginary-frequency grids~\cite{liu2016,Azizi2023,Azizi2024,Toelle2024}.
The precision of these numerical grids has undergone rigorous benchmarking against highly accurate $GW$ calculations across various systems, including solids~\cite{liu2016,Azizi2024}, two-dimensional materials~\cite{Azizi2024}, and molecules~\cite{foerster2021gw100,wilhelm2021,Azizi2024}. 
Overall, the findings indicate excellent numerical precision, typically better than 10\,meV for QP energies of the highest valence states and the lowest empty states~\cite{Azizi2024,wilhelm2021}.

A notable challenge remains in achieving high precision for five molecules within the \gwoh test set~\cite{gw100}, namely \ce{O3}, \ce{BeO}, \ce{MgO}, \ce{BN}, and \ce{CuCN}~\cite{wilhelm2021}, at the single-shot $G_0W_0$ level of theory~\cite{Golze2019} using the Perdew-Burke-Ernzerhof (PBE) functional~\cite{Perdew1996} for the underlying Kohn-Sham density functional theory~\cite{Kohn1965} (KS-DFT) calculation. We refer to this procedure as $G_0W_0@$PBE. 
For these molecules, the $G_0W_0$@PBE energies computed from low-scaling algorithms can differ by several hundred meV from reference calculations~\cite{wilhelm2021}.
In this work, we revisit these five molecules and we aim to demonstrate that the \reviewnew{multiple} $G_0W_0@$PBE solution\reviewnew{s of the QP equation are} \reviewout{is} unphysical and that a \reviewout{physical} \reviewnew{physically sound, single} solution can be obtained using partial eigenvalue self-consistency in the Green's function or using a hybrid functional as starting point for $G_0W_0$. 
We also aim to show that the low-scaling $GW$ algorithms are numerically precise for the $GW$ flavors which yield a \reviewout{physical} \reviewnew{single} solution. 
The article is organized as follows: 
We discuss the $G_0W_0$ scheme and multi-pole artifacts in the \gwoh    benchmark in Sec.~\ref{sec2a}. 
We describe the low-scaling space-time \textit{GW} algorithm in Sec.~\ref{sec2} and the reference $GW$ algorithm based on contour deformation in Sec.~\ref{sec3}. 
%
%
Computational details of our $GW$ calculations are given in Sec.~\ref{sec5}.
We present and discuss $G_0W_0$@PBE, \evgw and $G_0W_0$@PBE0 calculations in Sec.~\ref{sec6},~\ref{sec7} and \ref{sec:g0w0atpbe0}, respectively.
%

\section{Multisolution artifacts at the \textit{G}$_\text{0}$\textit{W}$_\text{0}$ level}
\label{sec2a}
In this section, we briefly introduce the $G_0W_0$ approach and different methods to obtain the QP solution, including the evaluation of the spectral function. We showcase the issue of multi\reviewout{-pole}\reviewnew{solution} artifacts for the MgO gas phase molecule, which is part of the \gwoh benchmark set~\cite{gw100}.
The \gwoh set is the standard molecular  test set for assessing the accuracy of $GW$ approaches and  $GW$ implementations~\cite{caruso2016, foerster2021gw100, maggio2017gw100, duchemin2021,wilhelm2021, wilhelm2018,govoni2018, bintrim2021,scott2023,monino2023,duchemin2020robust,mejia2021,wangy2021,Gao2019}.
\gwoh contains one hundred small molecules with covalent and ionic bonds covering a wide range of the periodic table.
%
%
The original \gwoh study~\cite{gw100} reports the $G_0W_0@$PBE QP energies of the highest occupied molecular orbital (HOMO) and the lowest occupied molecular orbital (LUMO) for all one hundred molecules. Those $G_0W_0@$PBE data are the ones against which new low-scaling algorithms are typically benchmarked~\cite{foerster2021gw100,wilhelm2021,Azizi2024}.

A $G_0W_0$ calculation starts from a self-consistent KS-DFT calculation~\cite{Kohn1965},
\begin{equation}
[h_0 (\br) + v_{\tmop{xc}} (\br)] \psi_n (\br) =
   \varepsilon_n^{\tmop{DFT}} \psi_n (\br)\,.
\end{equation}
$h_0$ contains the kinetic energy, the Hartree potential and
the external potential, while the exchange-correlation potential~$v_{\tmop{xc}}$ contains all electron-electron interactions beyond Hartree. $\psi_n(\br)$ is the KS orbital~$n$ and $\varepsilon_n^{\tmop{DFT}}$ the associated KS eigenvalue. 
The terms $G_0$ and $W_0$ indicate that the Green's function~$G$ and the screened Coulomb interaction~$W$ are both computed using KS orbitals and KS eigenvalues, i.e., self-consistent updates of $G$ and $W$ from Green's function theory are omitted in $G_0W_0$. 
A central object in a $G_0W_0$ calculation is the $G_0W_0$ correlation self-energy which can be expressed as~\cite{Golze2019,van2013gw}
\begin{equation}
 \Sigma_n^\text{c}(\omega) = \sum_m \sum_{s} \frac{\braket{\psi_n\psi_m|P_s|\psi_m\psi_n}}{\omega -\varepsilon_m^\text{DFT} + 
(\Omega_s^{\reviewnew{\text{RPA}}}-i\eta)\sgn(\varepsilon_{\mathrm{F}} -\varepsilon_m^\text{DFT})}\;.
    \label{e2}
\end{equation}
Here, $\omega$ denotes a frequency, $\varepsilon_\text{F}$  the Fermi level of KS-DFT and $\eta>0$ a broadening.
$\Omega_s^{\reviewnew{\text{RPA}}}$ with $\Omega_1^{\reviewnew{\text{RPA}}}\letext\Omega_2^{\reviewnew{\text{RPA}}}\letext\ldots$ are the charge-neutral singlet excitation energies  computed using the random phase approximation (RPA), i.e.~a vanishing  exchange-correlation kernel~\cite{Ullrich} \reviewnew{(see also discussion in Appendix~\ref{appb:Casida})}; $P_s$~represents the  product of transition densities of the excitation~$s$.~\cite{Golze2019,van2013gw}
One observable computed in a $G_0W_0$ calculation is the molecular spectral function, which takes the form of a many-body density of states~\cite{Onida2002,Golze2019},
\begin{align}
A(\omega) = \frac{1}{\pi}\,\text{Im} \; G(\omega)\;\text{sgn}(\varepsilon_\text{F}-\omega)\,,
\end{align}
where $G(\omega)$ is the trace over spatial arguments of the single-particle Green's function.
We express the spectral \mbox{function as}
\begin{align}
\label{spectral}
 A (\omega) &=  \sum_n\, A_n (\omega)\,,
\end{align}
where $n$ runs over all KS orbitals with a contribution $A_n(\omega)$ to the spectral function~\cite{Golze2020, Golze2019,marie2024gw,liu2016}
 \begin{align} \label{An}
 A_n (\omega) &= \frac{1}{\pi}\,\text{Im}\,\frac{\text{sgn}(\omega - \varepsilon_\text{F})}{\omega -
  (\varepsilon_n^{\tmop{DFT}} + \Sigma^\text{c}_n (\omega) + i\eta + \Sigma^\text{x}_n  -
  v_n^{\tmop{xc}})} 
  \\[0.5em]&= \label{An2}
  \frac{1}{\pi}\frac{ |\gamma| }{\left(\omega -
   (\varepsilon_n^{\tmop{DFT}} + \tmop{Re} \Sigma_n^\text{c} (\omega) +  \Sigma_n^\text{x} -
   v_n^{\tmop{xc}})\right)^2 + \gamma^2} \; .
\end{align}
Here, $\gamma\eqt \text{Im}\,\Sigma_n^\text{c}(\omega)\pt\eta $,  $\Sigma^\text{x}(\br,\br')$ is the exchange self-energy~\cite{Golze2019} and  $ \Sigma_n^\text{x}$ and $v_n^{\tmop{xc}}$ are the $n,n$-diagonal elements of the respective quantities.
The peak positions $\varepsilon_n^{G_0 W_0}$ of $A(\omega)$, known as QP energies, correspond to \reviewnew{the negative of the} ionization energies and \reviewnew{to} electron affinities of the molecule.
QP energies are the frequencies where the real part of the denominator in Eq.~\eqref{An} equals zero, i.e., the QP energies~$\varepsilon_n^{G_0 W_0}$  satisfy the QP equation
\begin{equation}
\label{energy-it}
\varepsilon_n^{G_0 W_0} = \varepsilon_n^{\tmop{DFT}} + \tmop{Re} \Sigma_n^\text{c}
   (\varepsilon_n^{G_0 W_0}) + \Sigma_n^\text{x} - v_n^{\tmop{xc}}\;.
\end{equation}
The non-linearity of $\Sigma_n^\text{c}(\omega)$ with respect to $\omega$, allows for  multiple solutions $\varepsilon_n^{G_0 W_0}$ for a given level~$n$.
Such multiple solutions have been observed for five molecules in the \gwoh test, namely O$_3$, BeO, MgO, BN, and CuCN at the $G_0W_0@$PBE level \cite{gw100}. 

\begin{figure}[]
    \centering
    \includegraphics[width=0.49\textwidth]{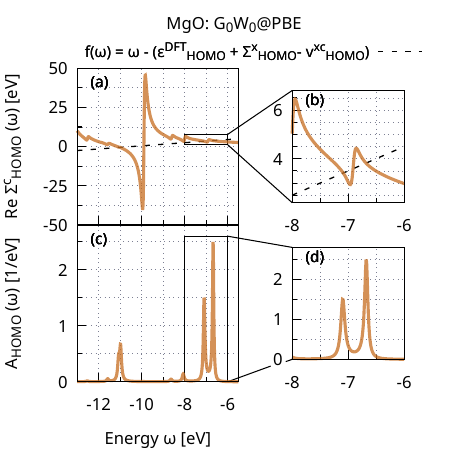}
    \caption{(a)/(b) Real part of the $G_0W_0$@PBE HOMO self-energy $\Sigma_\text{HOMO}^\text{c}(\omega)$, Eq.~\eqref{e2}, for the MgO molecule from the \gwoh set. 
    Intersections of the straight line $f(\omega)$ and Re\,$\Sigma_\text{HOMO}^\text{c}(\omega)$ lead to five solutions of the QP equation~\eqref{energy-it}, as already discussed in Ref.~\onlinecite{gw100}.
    (c)/(d) $G_0W_0$@PBE HOMO contribution~$A_\text{HOMO}(\omega)$ to the spectral function, Eq.~\eqref{An}.
    We have computed all quantities using the numerically precise contour deformation as implemented in FHI-aims~\cite{Golze2018} and a broadening~$\eta\eqt1$\,mHa, see further details in Sec.~\ref{sec3} and~\ref{sec5}.
     }
    \label{f0}
\end{figure}

We illustrate multiple solutions of Eq.~\eqref{energy-it} for state $n \eqt \text{HOMO}$ of the MgO molecule, see Fig.~\ref{f0}\,(a).
The black  dashed traces visualize the linear function
\begin{align}\label{fw}
    f(\omega) = \omega - (\varepsilon^\text{DFT}_\text{HOMO} 
    +
    \Sigma_\text{HOMO}^\text{x}
    -
    v^\text{xc}_\text{HOMO}) 
\end{align}
such that the intersections~$\varepsilon$ of $f(\omega)$ and $\text{Re}\,\Sigma^\text{c}_\text{HOMO}(\omega)$, $f(\varepsilon)\eqt \text{Re}\,\Sigma^\text{c}_\text{HOMO}(\epsilon)$, are the solutions of the QP equation~\eqref{energy-it}.
Five intersections are apparent,  $\varepsilon\intext\{-11.00\,\text{eV},-9.90\,\text{eV},-7.09\,\text{eV},-6.91\,\text{eV},-6.68\,\text{eV} \}$, where the highest solution has been picked in the original work~\cite{gw100} as $G_0W_0$@PBE HOMO energy.
The large number of intersections is due to the many poles in the self-energy~\eqref{e2}.

Each intersection~$\epsilon$ comes with a spectral weight~$Z_\varepsilon$ that quantifies the number of electrons associated with a given peak in the spectral function.
$Z_\varepsilon$ is usually approximately computed as
\begin{align}
 Z_\varepsilon\approx \left[1-\text{Re}
  \left. \frac{\partial \Sigma^\text{c}_n(\omega)}{\partial\omega }\right|_{ \displaystyle\omega=\varepsilon}\right]^{-1}\,,\label{qpweight}
\end{align}
see Appendix~\ref{sec:Z} for a derivation. 
Due to the small slope of $\Sigma^\text{c}_\text{HOMO}(\omega)$ at the intersections $\varepsilon\intext\{-7.09\,\text{eV}, -6.68\,\text{eV}\}$, the associated  spectral weight is large, whereas the other intersections exhibit a relatively small spectral weight.
This is also apparent in the HOMO contribution~$A_\text{HOMO}(\omega)$ to the $G_0W_0$@PBE spectral function~\eqref{An},  see Fig.~\ref{f0}\,(c)/(d). 
Three major peaks are visible \reviewnew{at --\,11.0~eV, --\,7.09~eV and --\,6.68~eV}, precisely located at intersections of $f(\omega)$ and $\text{Re}\,\Sigma^\text{c}_\text{HOMO}(\omega)$. 
\reviewnew{In particular, the two peaks at $\varepsilon\intext\{-7.09\,\text{eV}, -6.68\,\text{eV}\}$ carry both significant spectral weight, }
\reviewout{More precisely, $A_\text{HOMO}(\omega)$, Eq.~\eqref{An},  features two peaks at energy $\varepsilon\intext\{-7.09\,\text{eV}, -6.68\,\text{eV}\}$ with significant spectral weight,}
i.e.~a \reviewnew{distinct} single QP peak is absent.

\setlength{\tabcolsep}{5pt}
\def\arraystretch{1.5}
\begin{table}[tb!]
    \centering
        \caption{
   \reviewnew{ PBE and PBE0 eigenvalues and excitation energies $\Omega_s^\text{TDDFT/RPA}$ from time-dependent DFT (TDDFT) and RPA of the MgO molecule. All values in eV. In Appendix~\ref{appb:Casida}, we give details on TDDFT and RPA and we discuss the origin of the relationship $\Omega_1^\text{PBE}\apt\Omega_1^\text{PBE0}\apt\Omega_1^\text{RPA@PBE}\,{\ll}\,\Omega_1^\text{RPA@PBE0}$. TDDFT and RPA excitation energies have been computed using pyscf~\cite{Sun2020}.}}
    \begin{tabular}{lrr}
    \hline
    & \hspace{0.9cm}PBE\hspace{0.7cm}  & \hspace{0.9cm}PBE0\hspace{0.5cm} 
    \\
\hline
$\varepsilon_\text{HOMO--3}^\text{DFT}$ &
--\,19.52 & --\,22.13 
\\
$\varepsilon_\text{HOMO--2}^\text{DFT}$ &
--\,6.09 & --\,6.87 
\\
$\varepsilon_\text{HOMO}^\text{DFT} =\varepsilon_\text{HOMO--1}^\text{DFT} $& 
--\,4.79 & --\,6.07 
\\
$\varepsilon_\text{LUMO}^\text{DFT}$ & 
--\,4.29 & --\,3.60
\\
$\Omega_{s=1}^\text{TDDFT}=\Omega_{s=2}^\text{TDDFT}$ & 0.67 & 0.62 
 \\
$\Omega_{s=1}^\text{RPA}=\Omega_{s=2}^\text{RPA}$ & 0.82 & 2.83
 \\
$\Omega_{s=3}^\text{TDDFT}$ & 2.87 & 2.86 \\
$\Omega_{s=3}^\text{RPA}$ & 3.21 & 4.91 \\
\hline
    \end{tabular}

    \label{tab:PBEeigexc}
\end{table}
\reviewnew{
A closer inspection of $\Sigma^\text{c}_\text{HOMO}(\omega)$ at the two intersections   $\varepsilon\intext\{-7.09\,\text{eV}, -6.68\,\text{eV}\}$ in Fig.~\ref{f0}\,(b) reveals  a shallow pole of $\text{Re}\Sigma^\text{c}_\text{HOMO}(\omega)$ at $\omega_\text{pole}^\text{MgO}={-}\,6.91$\,eV.
As evident from Eq.~\eqref{e2}, poles in the real part of the self-energy  occur at 
\begin{align}
\omega_\text{pole} = \epsilon_i^\text{DFT}-\Omega_s^{\text{RPA}} 
 \hspace{1.0em}
 \text{or}
 \hspace{1.0em}
 \omega_\text{pole} = \epsilon_a^\text{DFT}+\Omega_s^{\text{RPA}}   \,,
 \label{polesSigma}
\end{align}
where $i$ denotes an occupied and $a$ an empty KS orbital index.
For a calculation with 2$N_\text{el}$ electrons and $N_\text{basis}$ basis functions, this makes in total $N_s\eqt N_\text{el}(N_\text{basis}{-} N_\text{el})$ excitation energies~$\Omega_s^{\reviewnew{\text{RPA}}}, s\eqt1,\ldots,N_s$ and thus $N_\text{basis}N_s$ poles in $\Sigma^\text{c}_\text{HOMO}(\omega)$~\cite{Veril2018}. 
We list KS eigenvalues and RPA charge-neutral excitation energies $\Omega_s^\text{RPA}$ of MgO in Table~\ref{tab:PBEeigexc} and we identify the self-energy pole at $-6.91$~eV based on Table~\ref{tab:PBEeigexc} as
\begin{align} 
\omega_\text{pole}^\text{MgO}  =
 \varepsilon_\text{HOMO--2}^\text{PBE} - \Omega_{s=1}^\text{RPA@PBE}
 = {-}6.91\;\text{eV}
\,.\label{e11}
\end{align}
%

Poles in the real-part of the self-energy have corresponding peaks in the imaginary part of $\Sigma^\text{c}$~\cite{Golze2019}. A distinct QP peak in $A_n(\omega)$ at~$\omega\eqt\varepsilon^*$ emerges when $\text{Im}\,\Sigma^\text{c}$ is small, as obvious from Eq.~\eqref{An2}. This is the case when the QP peak is far away from any pole of $\text{Re}\,\Sigma_\text{HOMO}^\text{c}(\omega)$. The spectral weight~$Z_{\varepsilon^*}$ of the QP peak is then close to one because of a small slope of the self-energy, \mbox{$| \text{Re}\,\partial_\omega\Sigma^\text{c}_n(\omega=\varepsilon^*)|\,{\ll}\,1$}~\cite{Golze2019}.  Further intersections of $f(\omega)$ and $\text{Re}\,\Sigma_\text{HOMO}^\text{c}(\omega)$ exist at energies lower than $\omega=\epsilon^*$. 
All these solutions are close to a pole in $\text{Re}\,\Sigma_\text{HOMO}^\text{c}(\omega)$. $\text{Im}\,\Sigma^\text{c}$ is consequently large, resulting in low spectral weights. In other words, we obtain further solutions with satellite character. In the molecular case, satellites are due to multielectron excitations, i.e., the charged electron or hole excitation couples to charge-neutral electronic excitations. 

The PBE eigenvalues $\epsilon_i^\text{PBE}$ of occupied KS states  are overestimated and usually a few electron volts higher in energy compared to the $GW$ energies. The poles at $\epsilon^{\text{PBE}}_i\mt\Omega_s^{\text{RPA}}$ are therefore at too high frequencies. Or in other words, the satellites occur erroneously at too high energy and are too close to the QP peak. For HOMO excitations computed at the $G_0W_0@$PBE level, satellites and QP peak are still well separated for most molecules and solids. Exceptions are, for example, the five molecules we discuss in this work. As we demonstrated for MgO, $\omega_\text{pole}^\text{MgO}  =
 \varepsilon_\text{HOMO--2}^\text{PBE} - \Omega_{s=1}^\text{RPA}$ is located close to the QP solution. This results in an erroneous redistribution of spectral weight from the QP to the satellite solution, ultimately causing the observed multisolution behavior. The location of the satellites becomes increasingly incorrect for deeper excitations. While $\epsilon^{\text{PBE}}_i$ is overestimated by a few electron volts for HOMO excitations, this overestimation is significantly larger, around 20\,--\,30 eV for 1s excitations of second-row elements~\cite{Golze2018}. In these cases, the poles are drastically too high in energy and the satellites always overlap with the QP solution, leading to a pronounced multisolution scenario with no identifiable QP peak at the $G_0W_0@$PBE level~\cite{Golze2020,li2022benchmark}. This stands in stark contrast to experimental observations, where one sharp 1s main peak with very small satellites is observed in X-ray photoemission spectra of closed shell molecules~\cite{Siegbahn1969all,Schirmer1987,Rocco2008}.  
%
%
%
%
}

\reviewout{
MgO is a molecule with large optical gap in the order of several eVs such that shake-up processes cannot lead to such a small splitting of peaks in  $A_\text{HOMO}(\omega)$.
This renders the two close peaks with similar spectral weight in $A_\text{HOMO}(\omega)$  unphysical.  
A clear QP peak is also absent in the $G_0W_0$@PBE spectral function of the other four challenging molecules BeO, BN, O$_3$ and CuCN, see Fig.~\ref{fa1}  and~\ref{fa2}  in the appendix. 
 Also in these cases, a single QP peak is expected.}
\reviewnew{We observe similar multi-solution behaviour as described for MgO for the HOMO of BeO, BN, O$_3$ and CuCN from the \gwoh set, see Fig.~\ref{fa1}  and~\ref{fa2}  in the appendix.}
These multiple solutions \reviewnew{at the $G_0W_0$@PBE level} are not only unphysical \reviewnew{as discussed above}, but they also pose a numerical challenge to low-scaling $GW$ algorithms~\cite{wilhelm2018,wilhelm2021}. 
In this work, we address this issue. 

\reviewnew{
\section{Correction of multipole artifacts with eigenvalue-selfconsistency in \textit{G}}
\label{sec:correction_multipole_arti}
The crucial role of eigenvalue self-consistency in the Green's function (ev$G$) has long been recognized in the literature~\cite{Pollehn1998,Gatti2015,Zhou2015,Martin_Reining_Ceperley_2016,Veril2018,Golze2022,li2022benchmark} as essential for preventing an artificial transfer of spectral weight from the QP peak to the satellites and suppressing multisolution behavior. 
%
%
We denote this procedure as \evgw  where 
$W$ is fixed at the $G_0W_0$ level, while $G$ in Eq.~\eqref{sigma-ls} is recomputed from Eq.~\eqref{green_spacetime} using the QP energies~\eqref{energy-it}.  %
 This procedure is repeated until self-consistency in the QP energies $\varepsilon_m^{\text{QP}}$ is reached~\cite{Golze2019}. In \evgwnospace, the energies $\varepsilon_m^\text{DFT}$ in Eq.~\eqref{e2} are replaced by $\varepsilon_m^{\text{QP}}=\varepsilon_m^\text{DFT} + \Delta^{\text{QP}}_m$, where $\Delta^{\text{QP}}_m$ is the QP correction for state $m$. The poles of Re\,$\Sigma^c_n$ (Eq.~\eqref{e2}) in \evgw are consequently at 
\begin{align}
\omega_\text{pole}  &= \varepsilon_i^\text{DFT} + \Delta^{\text{QP}}_i-\Omega_s^{\text{RPA}} \quad \text{for occupied orbital $i$}\,, \label{polesSigmaevgwocc}\\
 \omega_\text{pole}  &= \varepsilon_a^\text{DFT}+ \Delta^{\text{QP}}_a + \Omega_s^{\text{RPA}}  \quad \text{for empty orbital $a$}\,.
 \label{polesSigmaevgwvirt}
\end{align}
When using the PBE functional in KS-DFT, $\Delta^{\text{QP}}_m$ is usually negative for occupied states and positive for the empty states. Compared to $G_0W_0@$PBE, poles at frequency below the Fermi level thus often shift to more negative frequencies and poles at frequencies above the Fermi level to more positive frequencies. We will demonstrate this for the MgO case in Section~\ref{sec7}. 

We want to assess now in which frequency ranges poles in $\text{Re}\,\Sigma^n_c(\omega)$ occur at the ev$GW_0$ level, following the discussion of Véril \textit{\textit{et al}.}~\cite{Veril2018}. In addition to requiring ev$G$, we also assume positive, non-zero singlet RPA excitation energies $\Omega_s^{\text{RPA}} >0$, which should be the case for weakly correlated systems. For every level~$n$ close to HOMO and LUMO with
\begin{align}
\varepsilon_n^\text{QP}\in 
[\varepsilon_\text{HOMO}^{\text{QP}}\mt\Omega_1^{\text{RPA}},\varepsilon_\text{LUMO}^{\text{QP}}\pt\Omega_1^{\text{RPA}}]\,, \label{e12a}
\end{align}
the denominator of $\Sigma^\text{c}_n(\omega)$ in Eq.~\eqref{e2} is strictly non-zero (even for $\eta\rightarrowtext0$) because Eq.~\eqref{e12a} implies
\begin{align*}
\begin{split}
&\omega-\varepsilon_i^{\text{QP}}+ \Omega_s^{\text{RPA}}> 0\hspace{0.95em}\text{for $\omega\eqt\varepsilon_n^\text{QP}$, occupied orbital $i$, all $s$},
\\[0.3em]
&\omega-\varepsilon_a^{\text{QP}}-\Omega_s^{\text{RPA}}<0\hspace{0.95em}\text{for $\omega\eqt\varepsilon_n^\text{QP}$, empty orbital~$a$, all $s$.}
\end{split} 
\end{align*}
Therefore, $\Sigma^\text{c}_n(\omega{=}\varepsilon_n^\text{QP})$ is never singular for ev$G$, $\Omega_s\,{>}\,0$, and $n$ with~\eqref{e12a}~\cite{Veril2018}. Poles of $\text{Re}\,\Sigma^\text{c}_n(\omega)$ only appear at energies 
\begin{align}
\omega_\text{pole}^{\text{ev}GW_0} \in   
[-\infty,\varepsilon_\text{HOMO}^{\text{QP}}\mt\Omega_1^{\text{RPA}}]
\cup
[\varepsilon_\text{LUMO}^\text{QP}\pt\Omega_1^{\text{RPA}},\infty]\,.\label{eq:evgw0pole}
\end{align}
When considering $n\eqt\text{HOMO}$ or $n\eqt\text{LUMO}$ in ev$G$ with $\Omega_s^{\text{RPA}}\,{>}\,0$, all poles $\omega_\text{pole}^{\text{ev}GW_0}$ of $\Sigma^\text{c}_\text{HOMO/LUMO}(\omega)$ are thus separated from $\omega\eqt\varepsilon_\text{HOMO/LUMO}^{\text{QP}}$ by the energy $ \Omega_1^{\text{RPA}}$. By design, multiple solutions will not occur at the ev$GW_0$ level for HOMO and LUMO, given that the system has a non-zero optical gap. Furthermore, it has also been shown that ev$GW_0$ suppresses multisolution behavior for deeper states, inhibiting erroneous transfer of spectral weights from QP peak to satellites~\cite{Zhou2015,Golze2020}. We note that Eqs.~\eqref{polesSigmaevgwocc}\,--\,\eqref{eq:evgw0pole} hold also for the ev$GW$ scheme, where eigenvalues are also iterated in~$W$.
%

A computationally less expensive alternative to ev$G$ is introducing the so-called Hedin shift~\cite{hedin1965new,Hedin1999,Pollehn1998,Martin_Reining_Ceperley_2016,Golze2019,Golze2022} in the denominator of Eq.~\eqref{e2}. The Hedin shift can be viewed as an approximation to ev$GW_0$, where the individual shifts $\Delta_m^{\text{QP}}$ in Eqs.~\eqref{polesSigmaevgwocc} and \eqref{polesSigmaevgwvirt} are approximated with a global shift $\Delta \text{H}$. Another alternative is to start the $G_0W_0$ calculation from a hybrid DFT starting point, such as PBE0. In $G_0W_0@$PBE0, the poles will effectively also shift to more negative frequencies below the Fermi level because $\epsilon_i^{\text{PBE0}}<\epsilon_i^{\text{PBE}}$ and $\Omega_s^{\text{RPA@PBE0}}>\Omega_s^{\text{RPA@PBE}}$, as shown in Table~\ref{tab:PBEeigexc}. We note that the latter was also discussed in literature before~\cite{govoni2018,Golze2020}.  
In this work, we aim to show that low-scaling $GW$ is numerically precise for computing frontier QP energies  from $GW$ schemes that are free of multipole artifacts.
}

\section{Low-scaling \textit{GW} space-time algorithm}\label{sec2}
 
Many low-scaling $GW$ algorithms~\cite{foerster2011,stochastic_gw,liu2016,Gao2018,wilhelm2018,duchemin2019,foerster2020,Kim2020,Kutepov2020,wilhelm2021,duchemin2021} build on the $GW$ space-time method~\cite{rojas_1995}.
In this work, we execute $GW$ calculations using the low-scaling  algorithm from Ref.~\onlinecite{wilhelm2021} which adapts the  space-time method for use with Gaussian basis functions.
In order to introduce the basic idea of the $GW$ space-time method, we use a generic formulation in this section for non-periodic systems projecting all quantities on real-space grids.
It is important to
note that this formulation differs from the original $GW$ space-time method~\cite{rojas_1995} where some
quantities were calculated using a plane-wave basis set.

KS orbitals and eigenvalues are used to calculate the single-particle Green's function in imaginary time,
\begin{equation}
\label{green_spacetime}
  G (\br, \br', i \tau) = \begin{cases}
    \;\;\;\, i \sum\limits_i^{\tmop{occ}} \psi_i (\br) \psi_i^\reviewnew{*} (\br') e^{
    -|(\varepsilon_i^\text{DFT}-\varepsilon_\text{F}) \tau|}, & \tau < 0\;,\\[1em]
    - i \sum\limits_a^{\tmop{virt}} \psi_a^\reviewnew{*} (\br) \psi_a (\br') e^{
    -|(\varepsilon_a^\text{DFT}-\varepsilon_\text{F}) \tau|}, & \tau > 0\;,
  \end{cases} 
\end{equation}
where the sum over the index $i$ runs over all occupied KS orbitals and the sum over the index $a$ over all virtual, i.e., empy KS orbitals. $\varepsilon_\text{F}$ is the Fermi level. \reviewnew{We note that for the non-relativistic molecular calculations conducted in this work, the KS orbitals are real-valued.}
The irreducible polarizability follows,
\begin{equation}
\chi^0 (\br, \br', i \tau) = - iG (\br,
   \br', i \tau) G (\br, \br', - i \tau)\;,
\end{equation}
which is then \reviewout{Fourier} transformed to imaginary frequency,
\begin{align}
\chi^0 (\br, \br', i\omega)
  = i \int\limits_{-\infty}^\infty 
  e^{-i\omega\tau}\;
  \chi^0 (\br, \br',i\tau)\;\mathd \tau\,.
\end{align}
\reviewnew{This transform can be  understood as Laplace transform followed by analytic continuation to the imaginary axis and effectively takes the form of a Fourier transform~\cite{rieger1999gw}.}
Next, the dielectric function $\epsilon$ can be calculated in imaginary frequency
from the irreducible polarizability,
\begin{equation} 
\epsilon (\br, \br', i \omega) = \delta (\br,
   \br') - \int \mathd \br'' v (\br,
   \br'') \chi^0 (\br'', \br', i \omega)\;, \label{eps}
\end{equation}
using the Dirac delta function~$\delta(\br,\br')$ and the Coulomb interaction $v (\br, \br')\eqt 1/|\br\mt\br'|$. The screened
Coulomb interaction can be computed from the inverse dielectric function,
\begin{equation} 
W (\br, \br', i \omega) = \int \mathd \br''\,
   \epsilon^{- 1} (\br, \br'', i \omega) \,v
   (\br'', \br')\;. \label{W}
\end{equation}
It is convenient in $GW$ implementations to split the screened interaction~$W$ into the bare Coulomb interaction~$v$ and the correlation part~$W^\text{c}$,
\begin{align}
   W^\text{c} (\br, \br', i \omega)  = W (\br, \br', i \omega) - v(\br, \br')\,.
\end{align}
In the space-time method, $W^\text{c}$ is required in imaginary time,
\begin{align}
  W^\text{c} (\br, \br', i\tau)
  = \frac{i}{2\pi} \int\limits_{-\infty}^\infty 
  e^{i\omega\tau}\;
   W^\text{c} (\br, \br',i\tau)\;\mathd \omega\,,
\end{align}
and the correlation self-energy is given as product of the  Green's function and the screened Coulomb interaction,
\begin{equation}
\label{sigma-ls}
 \Sigma^\text{c} (\br, \br', i \tau) = i\,G (\br,
   \br', i \tau) W^\text{c} (\br, \br', i \tau)\;.
\end{equation}
The self-energy is then transformed to imaginary frequency, 
\begin{align}
  \Sigma^\text{c} (\br, \br', i\omega)
  = i \int\limits_{-\infty}^\infty 
  e^{-i\omega\tau}\;
   \Sigma^\text{c} (\br, \br',i\tau)\;\mathd \tau\,,\label{Sigmactime}
\end{align}
and we calculate its $(n,n)$-diagonal element,
\begin{align}
    \Sigma_n^\text{c} (i\omega) &= \langle \psi_n| \Sigma^\text{c} (i\omega)|\psi_n\rangle \nonumber
    \\&= \int \mathd \br \,\mathd \br' \,\psi_n^\reviewnew{*}(\br)\, \Sigma^\text{c} (\br, \br', i\omega)\,\psi_n(\br')\,.
\end{align}
The self-energy is then analytically continued to real frequency, i.e.~we determine the fit parameters~$a_{j,n}$ and $b_{j,n}$  of an $N$-parameter Pad\'e approximant~$P_n(i\omega)$~\cite{Vidberg1977,gw100} to match $ \Sigma_n^\text{c} (i\omega)$ of QP level~$n$ as closely as possible,
\begin{align}
 \Sigma_n^\text{c} (i\omega) \approx  P_n(i\omega) = \frac{\displaystyle\sum\limits_{j=1}^{(N-1)/2}a_{j,n}\cdot (i\omega)^j}{\displaystyle1+\sum\limits_{j=1}^{N/2}b_{j,n}\cdot (i\omega)^j}\,.
\end{align}
Based on the identity theorem for analytic functions~\cite{ahlfors1979}, one evaluates~$P_n$ at real frequencies to compute the self-energy at real frequencies, $ \Sigma_n^\text{c} (\omega)\apt P_n(\omega)$~\cite{gw100,Golze2019}.

Focusing on the $G_0W_0$ method already introduced before, we use KS orbitals to  approximate the QP wavefunctions and we compute $G$ and $W$ only once using KS orbitals and KS eigenvalues from Eqs.~\eqref{green_spacetime}\,--\,\eqref{W}.
The QP energies~$\varepsilon_n^{G_0 W_0}$ can finally be calculated by solving the QP equation~\eqref{energy-it}.

\reviewout{
In this work, we also employ \evgw  where 
$W$ is fixed at the $G_0W_0$ level, while $G$ in Eq.~\eqref{sigma-ls} is recomputed from Eq.~\eqref{green_spacetime} using the QP energies~\eqref{energy-it}.
 This procedure is repeated until self-consistency  in the QP energies is reached~\cite{Golze2019}.}

%

The computational cost of the presented algorithm increases with $O(N^3)$ in the system size~$N$.
This is apparent from Eq.~\eqref{green_spacetime}, which requires
${O} (N^2_{\tmop{grid}}(N_{\tmop{occ}} {+} N_{\tmop{vir}})) \eqt
{O} (N^3)$ number of floating point operations, where $N_{\tmop{grid}}$ is the number of real-space grid points, $N_{\tmop{occ}}$ the number of occupied KS orbitals, and $N_{\tmop{vir}}$ the number of virtual KS orbitals.

In a standard $GW$ algorithm, a computational bottleneck is evaluating the Adler-Wiser formula~\cite{Adler1962,Wiser1963} for the 
irreducible polarizability in imaginary frequency,
\begin{equation}
\chi^0 (\br, \br', i \omega) = \sum_i^{\tmop{occ}} 
   \sum_a^{\tmop{virt}}  \frac{2 (\varepsilon_i -
   \varepsilon_a)}{(\varepsilon_i - \varepsilon_a)^2 + \omega^2} \psi_i
   (\br) \psi_a^\reviewnew{*} (\br) \psi_i^\reviewnew{*} (\br') \psi_a
   (\br')\,.
\end{equation}
This computation requires ${O} (N^2_{\tmop{grid}}
 N_{\tmop{occ}}N_{\tmop{virt}}) = {O} (N^4)$ operations, and is thus computationally more demanding than the whole low-scaling algorithm~\eqref{green_spacetime}\,--\,\eqref{Sigmactime} for a large system.


The drawback of algorithm~\eqref{green_spacetime}\,--\,\eqref{Sigmactime} is that it requires very fine real-space grids, in particular when evaluating Coulomb interactions in Eq.~\eqref{eps} and~\eqref{W}.
This is why the original $GW$ space-time method~\cite{rojas_1995} used a plane-waves basis for the calculation of $\varepsilon (i \omega)$ and $W (i \omega)$.
Then, the $O(N^3)$-scaling convolutions~\eqref{eps} and~\eqref{W} in real space transform into $O(N^2)$ multiplications in the plane-wave basis. 
The inversion of $\epsilon(i\omega)$ remains as cubic-scaling step. 
For the low-scaling $GW$ implementation used in this work~\cite{wilhelm2021}, Eq.~\eqref{green_spacetime}\,--\,\eqref{Sigmactime}  have been reformulated in a Gaussian basis set  resulting in  effective ${O} (N^2)$ scaling~\cite{wilhelm2018,wilhelm2021}.

\section{Reference \textit{GW} calculations with Contour Deformation}\label{sec3}

In the low-scaling $GW$ space-time method, one major challenge regarding numerical precision is the analytic continuation of $\Sigma^\text{c}(i\omega)$ to $\Sigma^\text{c}(\omega)$.
A numerically more precise procedure for computing $\Sigma^\text{c}(\omega)$ is the contour deformation (CD)~\cite{Godby1988,Lebegue2009,Gonze2009,Golze2018,Golze2019}.
The starting point of CD is Eq.~\eqref{sigma-ls}, formulated in real frequency,~\cite{Golze2019} 
\begin{align} 
\label{sigma-realspace}
    \Sigma^\text{c} &(\br,\br', \omega) = 
    \frac{i}{2 \pi} \int\limits_{-\infty}^\infty 
    G (\br,\br',\omega+\omega')\, W^\text{c} (\br,\br',\omega')\,\mathd \omega' \;.
\end{align}
Due to the pole structure of $G$ and $W$ on the real-frequency axis, the numerical integration of Eq.~\eqref{sigma-realspace} is potentially unstable~\cite{Golze2019}.
One way to circumvent this problem is to rewrite Eq.~\eqref{sigma-realspace} with a contour integral, 
\begin{align}
    \label{cd}
    \Sigma^\text{c}( \br, \br', \omega ) =& 
        \frac{i}{2 \pi} \oint \mathd \omega' \, G( \br, \br', \omega + \omega')\,W^\text{c}( \br, \br', \omega' ) \\
    &- \frac{1}{2 \pi} \int\limits_{-\infty}^\infty \mathd \omega' \, G( \br, \br', \omega + i \omega' )\, W^\text{c}( \br, \br', i \omega' ) \,,\nonumber
\end{align}
where the closed integral comprises the real axis, two arcs  and the imaginary axis.
The closed integral can be calculated using Cauchy's residue theorem, while the imaginary axis can be integrated numerically since the problematic pole structure of $G$ and $W$ is restricted to the real frequency axis.~\cite{Golze2019}
QP energies from CD follow from solving the QP equation~\eqref{energy-it}. 

The scaling for evaluating the self-energy from CD is $O(N^4)$ for a single valence excitation~\cite{Golze2018}, which 
 is an order higher than the $O(N^3)$-scaling of the $GW$ space-time method. 
 CD is numerically highly accurate. It was shown that CD reproduces the exact self-energy structure by comparing with fully analytic solutions obtained by evaluating Eq.~\eqref{e2}~\cite{Golze2018}. We use therefore the CD approach as reference to assess the numerical precision of our low-scaling $GW$ algorithm. 
More details on CD can be found in Refs.~\onlinecite{Golze2018,Golze2019}.


\section{Computational Details}\label{sec5}

\subsection{Low-scaling \textit{GW} calculations using CP2K}
The low-scaling $GW$ space-time algorithm used in this work~\cite{wilhelm2021} is implemented in the CP2K software package~\cite{Kuehne2020,cp2k}.
CP2K employs a Gaussian basis set for representing KS orbitals.  We use the Gaussian and augmented plane-waves  scheme (GAPW)~\cite{Lippert1999} in CP2K which allows for all-electron calculations. 
The low-scaling $G_0W_0$ implementation~\cite{wilhelm2021}  is a reformulation of the space-time method~\cite{rojas_1995} in a Gaussian basis set, where sparsity is introduced by combining a global resolution-of-the-identity (RI) scheme with a truncated Coulomb metric~\cite{Vahtras1993,Jung2005}.

For expanding KS orbitals, we use the def2-QZVP  basis set~\cite{weigend2003} with an RI-cc-pV5Z auxiliary basis~\cite{Haettig2005}.
We choose a minimax time-frequency grid~\cite{kaltak2014low,liu2016,Azizi2023,Azizi2024} with 30 points for all low-scaling $GW$ calculations.
For the RI with the truncated Coulomb metric we set a truncation radius  of~3\,\AA~\cite{Vahtras1993,Jung2005}.
We compute two- and three-center integrals over Gaussians using recursive, analytical schemes~\cite{Golze2017,libint}.  
The self-energy is analytically continued from imaginary frequency to the real frequency using a Pad\'{e} model~\cite{Vidberg1977,gw100} with 16 parameters.
We  broaden the spectral functions using  $\eta \eqt \SI{1}{mHa} \eqt \SI{27.2}{meV}$ in Eq.~\eqref{An}.

\subsection{Reference \textit{GW} calculations with contour deformation using FHI-aims}
We use the CD-$GW$ algorithm~\cite{Golze2018} as implemented in the FHI-aims software package~\cite{Blum2009}.
FHI-aims is an all-electron electronic structure code using numerical atom-centered orbitals (NAO) for expanding KS orbitals \cite{Blum2009}. 
We use the Gaussian def2-QZVP~\cite{weigend2003} basis set, which can be also represented as NAO basis,  and an automatically generated RI basis set. 
In the CD, we employ a modified Gauss-Legendre grid~\cite{ren2012} for the evaluation of the imaginary-frequency integral term, setting the number of frequency points to 2000 to obtain benchmark quality. 
The broadening parameter~\cite{Golze2018} is set to a rather large value $\eta \eqt \SI{4}{mHa} \eqt \SI{108.8}{meV}$ to facilitate the convergence in the ev$GW_0$ case.
The only exception is that for Fig.~\ref{f0}, we use a smaller broadening of 1\,mHa to obtain sharp and well-separated peaks in the spectral function.
The broadening of peaks in the spectral function resulting from CD differs from that of analytic continuation, as outlined in Ref.~\citenum{Golze2018}. 
Therefore it is difficult to directly compare peak widths and heights in the spectral function between CD and analytic continuation.

\section{Low-scaling \textit{G}$_\text{0}$\textit{W}$_\text{0}$@PBE calculations on \ce{O3}, B\MakeLowercase{e}O, M\MakeLowercase{g}O, \ce{BN}, C\MakeLowercase{u}CN}
\label{sec6}

In this section we present low-scaling $G_0W_0$@PBE calculations on the five  molecules, \ce{O3}, \ce{BeO}, \ce{MgO}, \ce{BN}, and \ce{CuCN}, and we will discuss the challenges that occurred.

It was already observed in the original \gwoh work~\cite{gw100} that computing the $G_0W_0$@PBE HOMO energy of \ce{O3}, \ce{BeO}, \ce{MgO}, \ce{BN}, and \ce{CuCN} posed a significant numerical challenge.
In the analytic continuation, this challenge was addressed by employing a 128-pole function to fit $\Sigma^\text{c}_\text{HOMO}(i\omega)$ for obtaining $\Sigma^\text{c}_\text{HOMO}(\omega)$ along the real-frequency axis. 
This high-precision fit was essential for accurately computing  the $G_0W_0$@PBE HOMO energy from the QP equation~\eqref{energy-it} for the five aforementioned molecules. 
For all other 95 molecules, a 16-pole fit on $\Sigma^\text{c}_\text{HOMO}$ was sufficient for accurately computing the $G_0W_0$@PBE HOMO energy. 
$GW$ space-time algorithms require in general three \reviewout{Fourier} transforms between imaginary time and imaginary frequency, as discussed in Sec.~\ref{sec2}.
These transforms are executed numerically using discrete time and frequency grids.
The functions in  imaginary time and frequency   usually have long tails and localized features. 
The usual Fast Fourier Transform with homogeneously spaced integration grids would need a large number of discrete time or frequency points. 
Instead, a non-equidistant Fourier transform can reduce the number of time and frequency points drastically~\cite{liu2016,Azizi2024}. 
Non-equidistant grids can be set up through various techniques aimed at identifying discrete  points that are optimal in a certain sense. Such approaches include the minimax approximation~\cite{takatsuka2008minimax,delben2015enabling,HELMICHPARIS2016,liu2016,wilhelm2018,wilhelm2021,hackbusch2019computation,Azizi2023,Azizi2024} or least-square quadratures~\cite{haeser1992,Ayala1999,Lambrecht2005,Kats2008,Kats2009,foerster2020}.
Both algorithms have in common that they are numerically ill-conditioned and grid generation is typically restricted to less than 100 grid points.
For instance, the recently introduced GreenX library~\cite{Azizi2023} offers minimax time and frequency grids, each containing up to 34 points~\cite{Greenx,Azizi2023,Azizi2024}.
Consequently, when utilizing time and frequency grids from the GreenX library, the number of parameters available for fitting $\Sigma^\text{c}(i\omega)$ is limited to 34.
This is not sufficient for accurately computing the $G_0W_0$@PBE HOMO energy of the five challenging molecules, \ce{O3}, \ce{BeO}, \ce{MgO}, \ce{BN}, \ce{CuCN},~\cite{wilhelm2018,wilhelm2021,Azizi2024} which required a fit with a 128-pole model~\cite{gw100}, that requires at least 128 imaginary-frequency points.
\begin{figure}[]
    \centering
    \includegraphics[width=0.49\textwidth]{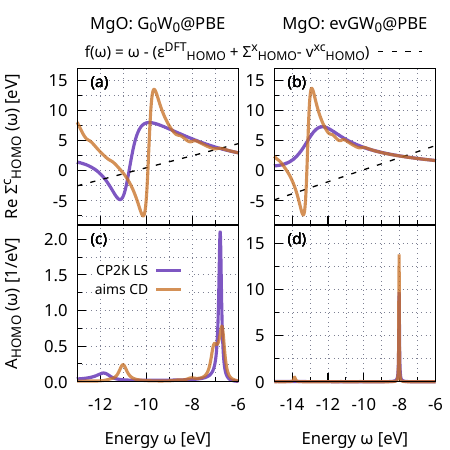}
    \caption{Real part of the HOMO self-energy $\Sigma_\text{HOMO}^\text{c}(\omega)$ [$G_0W_0$@PBE in (a) and \evgwnospace@PBE in (b)] and the HOMO contribution to the spectral function~$A_\text{HOMO}(\omega)$ [Eq.~\eqref{An}, $G_0W_0$@PBE in (c) and \evgwnospace@PBE in (d)] for the molecule \ce{MgO}; computed from low-scaling $GW$ in the CP2K package~\cite{wilhelm2021} ("CP2K LS"; violet traces) and from the highly accurate CD as implemented in FHI-aims~\cite{Golze2018} ("aims CD"; brown traces).
    $\text{Re}\Sigma_\text{HOMO}^\text{c}(\omega)$ and $A(\omega)$ for the other four numerically challenging molecules \ce{BeO}, \ce{O3}, \ce{BN}, CuCN from the \gwoh test set~\cite{gw100} are available in Figs.~\ref{fa2} and~\ref{fa3} for $G_0W_0$@PBE and \evgwnospace@PBE in the appendix.  
  }
    \label{f1}
\end{figure}

We now use our low-scaling $GW$ implementation~\cite{wilhelm2021}   to compute the $G_0W_0@$PBE self-energy and spectral function.
The results for the MgO molecule are shown as violet traces in Fig.~\ref{f1}\,(a) and~(c). 
Low-scaling $GW$~\cite{wilhelm2021} fails to reproduce the shallow self-energy pole at \reviewout{$\omega\apt{-}\,7.0\,$eV}\reviewnew{$\omega_\text{pole}\apt{-}\,6.91\,$eV} [Fig.~\ref{f1}\,(a) and zoom in Fig.~\ref{f0}\,(b)] which results in a single peak of the spectral function at $\omega\apt{-}\,6.9\,$eV [Fig.~\ref{f1}\,(c)] instead of the split peak present in the reference CD calculation [Fig.~\ref{f1}\,(c)]. 
The same failure in computing the self-energy has been observed in the original work, Ref.~\onlinecite{gw100}~(Fig.~13), when using analytical continuation with a 16-pole approximant to the self-energy.
When using a 128-pole approximant for analytical continuation all poles of the self-energy were correctly reproduced~\cite{gw100}.
The low-scaling $GW$ implementation~\cite{wilhelm2021} is currently restricted to at most 34 imaginary-frequency points, and thus 34 fit parameters for analytic continuation.
The Pad\'e model is thus not flexible enough to accurately reproduce the $G_0W_0$@PBE self-energy matrix elements for the HOMO of MgO. The same  observation is made for \ce{O3}, \ce{BeO},  \ce{BN}, and \ce{CuCN} (Fig.~\ref{fa2} in the appendix).
\reviewnew{
As discussed in Sec.~\ref{sec2a} and~\ref{sec:correction_multipole_arti}, the multiple solutions at the $G_0W_0@$PBE level of theory are an artifact caused by the lack of eigenvalue self-consistency in $G$. 
Therefore, we should disregard the unphysical multi-solution $G_0W_0@$PBE calculations and instead investigate whether we can  obtain numerically precise  \evgw HOMO QP energies using low-scaling $GW$.
}
\reviewout{As discussed in Sec.~\ref{sec2a} the multiple solutions at the $G_0W_0@$PBE level of theory are an artifact.
It was found that the erroneous multi-solution behavior at the $G_0W_0@$PBE level becomes increasingly pronounced with higher binding energies of the occupied states. Most notably, it has been observed that $G_0W_0@$PBE generally fails to reproduce a unique QP solution for molecular 1s core states.~\cite{Golze2020} The failure was traced back to an extreme transfer of spectral weight from the QP peak to the satellite spectrum. 
It was shown that the correct physics, a single QP peak for core states, can be restored by using eigenvalue self-consistent schemes, renormalized singles or by using $G_0W_0$ with hybrid functionals with almost 50\,\% of exact exchange as a starting point.}~\cite{Golze2020,li2022benchmark} \reviewout{The best quantitative results with respect to experiment were obtained with ev$GW_0$@PBE or a Hedin shift in the Green's function, which can be viewed as an approximation of ev$GW_0$.}~\cite{li2022benchmark}  
\begin{figure*}[]
    \centering\includegraphics[width=\textwidth]{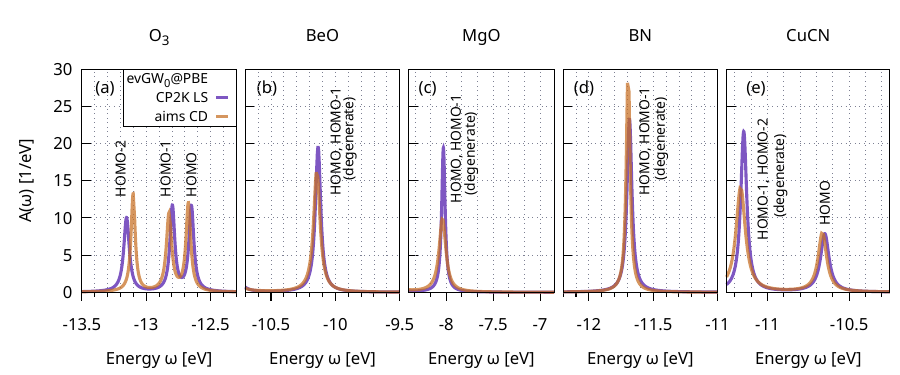}
    \caption{(a)-(e) ev$GW_0$@PBE spectral function  for five numerically challenging molecules (\ce{O3, BeO, MgO, BN, CuCN}) computed from low-scaling $GW$~\cite{wilhelm2021} (CP2K LS) and from CD~\cite{Golze2018} (aims CD).
    Peak positions and spectral weight of the HOMO peaks are listed in Table~\ref{tab:spectral}.
    $G_0W_0$@PBE and $G_0W_0$@PBE0 spectral functions  are shown in Fig.~\ref{fa1b} and~\ref{fa1}.
    Graphical solutions of the QP equation~\eqref{energy-it} are also shown in the appendix,  Fig.~\ref{fa2}, \ref{fa3} and~\ref{fa4} for $G_0W_0$@PBE, \evgwnospace@PBE, and $G_0W_0$@PBE0, respectively.
    }
    \label{f2}
\end{figure*}
\section{Numerically precise low-scaling  \lowercase{ev}\textit{G}\textit{W}$_\text{0}$ calculations on \ce{O3}, B\MakeLowercase{e}O, M\MakeLowercase{g}O, \ce{BN}, C\MakeLowercase{u}CN}
\label{sec7}
\reviewout{Motivated by the studies on  core levels,~\cite{Golze2018,Golze2020,li2022benchmark} we explore \evgw aiming to restore proper QP peaks in the spectral function of the five molecules \ce{O3}, \ce{BeO}, \ce{MgO}, \ce{BN}, \ce{CuCN}.}
 \reviewnew{Following the discussion in Sec.~\ref{sec:correction_multipole_arti}, \evgwnospace@PBE ensures that poles of the self-energy are separated from  frontier QP energies. 
 This cures the unphysical and numerically challenging multi-solution issue of the HOMO at the $G_0W_0$@PBE level and we thus} 
 explore \evgw to calculate the spectral function of the five molecules \ce{O3}, \ce{BeO}, \ce{MgO}, \ce{BN}, \ce{CuCN}.
%
%
%
%
The spectral function $A_{\text{HOMO}}$ for the MgO molecule at the \evgwnospace@PBE level is shown in Fig.~\ref{f1}\,(d). 
Only a single QP peak in the spectral function is present at $\omega\eqt8.03$\,eV.
The QP peak computed from the low-scaling algorithm~\cite{wilhelm2021} and the CD~\cite{Golze2018} match in position (within 9 meV) and spectral weight (within 1\,\%, Table~\ref{tab:spectral}). 
We investigate the drastic difference between the spectral function computed with $G_0W_0$  and \evgw by inspecting the respective self-energies, Fig.~\ref{f1}\,(a) and~(b).
For the \evgw self-energy [Fig.~\ref{f1}\,(b)] the poles are shifted by $\sim$\,2\,eV towards lower energy compared to the $G_0W_0$ self-energy [Fig.~\ref{f1}\,(a)].
\reviewnew{The \evgw self-energy does not exhibit  poles at the HOMO QP energy of 8.03\,eV [Fig.~\ref{f1}\,(b)].
This is expected, as for \evgwnospace, self-energy poles are separated from the HOMO QP energy by at least the lowest RPA excitation energy~$\Omega_1^{\text{RPA}}$, as discussed in Sec.~\ref{sec:correction_multipole_arti}.
}
The consequence is that there is only a single crossing point of the self-energy and the dashed straight line $f(\omega)$, Eq.~\eqref{fw}, close to $\omega\eqt8.03$\,eV.
The slope of the self-energy is small at this crossing point such that the associated spectral weight is close to one. 
The same is observed for the other four molecules (Fig.~\ref{fa3} in the appendix).

\setlength{\tabcolsep}{5pt}
\def\arraystretch{1.4}
\begin{table}[t!]
    \centering\vspace{-1em}
        \caption{ \evgwnospace@PBE HOMO energy, $\varepsilon^{\text{ev}GW_0\text{@PBE}}_\text{HOMO}$ and spectral weight~$Z_\text{HOMO}$ of the HOMO peak, computed from  CD~\cite{Golze2018} and from low-scaling (LS) $GW$~\cite{wilhelm2021} for the five multi-solution cases from the \gwoh set using a def2-QZVP basis set. 
    The spectral weight $Z$ has been computed according to Eq.~\eqref{lorentz}.
    \reviewnew{As comparison, we also show the $G_0W_0$@PBE HOMO energy~\cite{gw100}, the CCSDT(T) ionization potential~\fbout{\cite{Krause2015}}\fbnew{\cite{bruneval2021gw}} (computed using a  def2-TZVPP basis set) and the  experimental  vertical ionization potential (IP).}
    }
    \begin{tabular}{  l | *{5}{>{\centering\arraybackslash}p{23pt}} }
    \hline
          &\ce{O3} &\ce{BeO} &\ce{MgO} &\ce{BN} & \ce{CuCN}\\\hline   
   $|\varepsilon^{\text{ev}GW_0\text{@PBE}}_\text{HOMO}|$ (CD)& 12.667 & 10.141 & 8.039 & 11.695  & 10.663 \\
        $|\varepsilon^{\text{ev}GW_0\text{@PBE}}_\text{HOMO}|$ (LS)  & 12.649 & 10.134 & 8.030 & 11.685 & 10.649 \\
         $Z_\text{HOMO}$ (CD) & 0.83 & 0.85 & 0.68 &  0.85 & 0.79 \\
         $Z_\text{HOMO}$ (LS) & 0.88 & 0.78 & 0.68 & 0.85 & 0.80\\
\reviewnew{$|\varepsilon^{G_0W_0\text{@PBE}} _\text{HOMO}|$ \cite{gw100}} &11.39 & 8.62  &6.66 & 11.01 & 9.42
\\
\reviewnew{CCSD(T)\fbout{~\cite{Krause2015}}\fbnew{~\cite{bruneval2021gw}}} & \fbout{12.55}\fbnew{12.71}
& 9.94
& \fbout{7.49}\fbnew{7.91}
& \fbout{11.89}\fbnew{11.98}
&  10.85  \\
\reviewnew{\text{IP (exp.)}}&12.73~\cite{gw100}\;&10.10~\cite{gw100}&\;8.01\cite{Bellert2000}&\;\;--&--\\
\hline
    \end{tabular}
    \label{tab:spectral}
\end{table}

We report the total spectral function $A(\omega)$ for all five challenging molecules,  \ce{O3}, \ce{BeO}, \ce{MgO}, \ce{BN}, \ce{CuCN}, from \evgw in Fig.~\ref{f2}.
For all molecules, the HOMO peak positions of low-scaling $GW$~\cite{wilhelm2021} and CD~\cite{Golze2018} agree well, the mean absolute deviation is only 12\,meV (Table~\ref{tab:spectral}).
The width of the QP peaks differ between low-scaling $GW$ and the CD; the reason is the inherently different definition of the broadening parameter~$\eta$ in the two algorithms, cf.~Eq.~\eqref{An} and Ref.~\citenum{Golze2018}.
Our work shows that low-scaling space-time \evgwnospace@PBE is numerically precise for \ce{O3}, \ce{BeO}, \ce{MgO}, \ce{BN}, \ce{CuCN}. 
The improvement compared to the $G_0W_0$@PBE is due to the fact that ev$GW_0$@PBE yields well-defined  QP valence peaks, whereas $G_0W_0@$PBE \reviewout{fails to produce a physical result} \reviewnew{exhibits unphysical, erroneous multi-solution behaviour as discussed in Sec.~\ref{sec2a} and~\ref{sec:correction_multipole_arti}}. 
\reviewnew{The failure of $G_0W_0$@PBE also reflects in Table~\ref{tab:spectral}: \evgwnospace@PBE HOMO energies show significantly better agreement with   CCSD(T) calculations and experimental ionization potentials compared to $G_0W_0$@PBE HOMO energies.}
Therefore, it is not reasonable to invest effort in developing low-scaling $GW$ methods which can recover the $G_0W_0$@PBE solution for these five molecules, as it is an insufficient level of theory from the outset. \reviewnew{It was also shown for solids that \evgwnospace@PBE outperforms $G_0W_0$@PBE for computing accurate bandgaps, see discussion in Refs.~\cite{Shishkin2007} and~\cite[p.~29]{Golze2019}.
\evgwnospace@PBE is thus suitable for both molecules and solids making it ideal for studying complex, disordered systems such as molecules on surfaces or interfaces. 
In case of issues related to PBE orbitals used in \evgwnospace@PBE, one might also refer to \evgwnospace@HSE06~\cite{Heyd2003}\footnote{{In case of issues with HSE06 orbitals, one might consider  quasiparticle-selfconsistent $GW$ (qs$GW$)~\cite{Schilfgaarde2006}.}} for complex systems. 
}
\reviewout{The  enhanced computational cost of \evgw over $G_0W_0$  might be reduced again by using Hedin shifts}~\cite{Pollehn1998,li2022benchmark}.

\reviewnew{
Another way of partial self-consistency is the ev$GW$ scheme. 
Our low-scaling algorithm is also expected to reproduce ev$GW$@PBE results well since the multi-solution issue is resolved by  eigenvalue self-consistency in $G$. However, ev$GW$@PBE tends to underscreen, a phenomenon that is less noticeable for molecular frontier orbitals but becomes more apparent for deeper states~\cite{li2022benchmark}. This underscreening is particularly evident in solids, where it has been shown that ev$GW$@PBE produces bandgaps that are too large compared to experimental values~\cite{Shishkin2007,Golze2019}. The effectiveness of \evgwnospace@PBE arises from a fortunate error cancellation: while the PBE approximation typically underestimates the fundamental gap, using PBE eigenvalues in $W$ leads to an overscreened potential. This overscreening in $W$ counteracts the underscreening caused by the absence of vertex corrections, which is another way of expressing the analysis in Appendix~\ref{appb:Casida}. 
%
%
%
}

\section{Numerically precise low-scaling  \textit{G}$_\text{0}$\textit{W}$_\text{0}@$PBE0 calculations on \ce{O3}, B\MakeLowercase{e}O, M\MakeLowercase{g}O, \ce{BN}, C\MakeLowercase{u}CN}
\label{sec:g0w0atpbe0}
\setlength{\tabcolsep}{5pt}
\def\arraystretch{1.5}
\begin{table}[b!]\vspace{-1em}
    \centering
    \caption{
    $G_0W_0$@PBE0 HOMO energy, $\varepsilon^{G_0W_0\text{@PBE0}}_\text{HOMO}$ and spectral weight~$Z_\text{HOMO}$ of the HOMO peak, computed from CD~\cite{Golze2018} and from low-scaling (LS) $GW$~\cite{wilhelm2021} for all five difficult molecules from the \gwoh set using a def2-QZVP basis set.
     The spectral weight $Z$ has been computed according to Eq.~\eqref{lorentz}.
    }
    \begin{tabular}{  l | *{5}{>{\centering\arraybackslash}p{23pt}} }
    \hline
          &\ce{O3} &\ce{BeO} &\ce{MgO} &\ce{BN} & \ce{CuCN} \\\hline
         $|\varepsilon^{G_0W_0\text{@PBE0}}_\text{HOMO}|$ (CD)& 12.566 & 9.646 & 7.459 & 11.545 & 10.253 \\
        $|\varepsilon^{G_0W_0\text{@PBE0}}_\text{HOMO}|$ (LS)  & 12.557 & 9.643 & 7.454 & 11.542 & 10.230 \\
         $Z_\text{HOMO}$ (CD) & 0.81 & 0.76 & 0.67 & 0.83 & 0.81 \\
         $Z_\text{HOMO}$ (LS) & 0.81 & 0.77 & 0.67 & 0.83 & 0.81 \\\hline
    \end{tabular}
    
    \label{tab:pbe0}
\end{table}
\begin{figure*}
    \centering
\vspace{-1em}
\includegraphics[width=\textwidth]{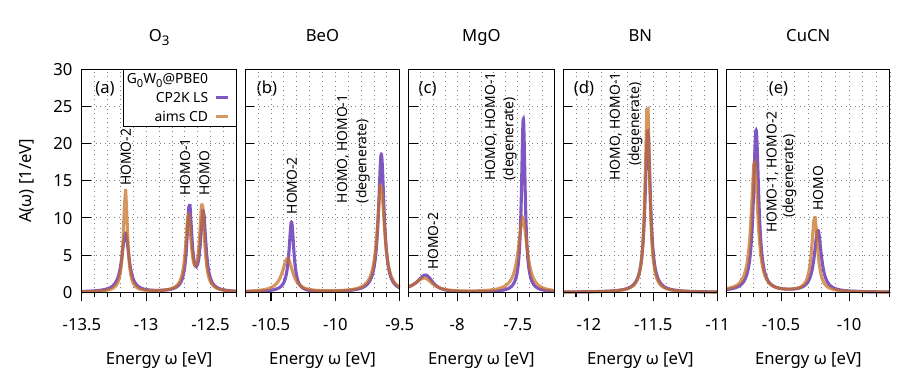}
    \caption{(a)-(e) $G_0W_0$@PBE0 spectral function  for \ce{O3}, \ce{BeO}, \ce{MgO}, \ce{BN}, \ce{CuCN} computed from low-scaling $GW$ in CP2K~\cite{wilhelm2021} (CP2K LS) and from CD in FHI-aims~\cite{Golze2018} (aims CD).}
    \label{fa1b}
\end{figure*}

\reviewout{We also compute} \reviewnew{The} HOMO QP energies at the $G_0W_0$@PBE0 level \reviewnew{are given in} Table~\ref{tab:pbe0}.
The low-scaling algorithm reproduces the QP energies from CD  within 9 meV and the spectral weights within 0.01. 
The total spectral functions $A(\omega)$ displayed in Fig.~\ref{fa1} follows also closely the CD reference. 
This alignment occurs because  
 the poles in the real part of $\Sigma^\text{c}_{\text{HOMO}}$ shift by 2\,--\,3\,eV towards more negative frequencies (Fig.~\ref{fa4} in the appendix), similar to what is observed in ev$GW_0$@PBE calculations. This shift is attributed to the the fact that \reviewout{$\epsilon_{\text{HOMO}}^\text{DFT}$} \reviewnew{$\epsilon_{i}^\text{DFT}$ ($i$: occ.~index)}  are more negative at the PBE0 than the PBE level, \reviewnew{and $\Omega_s^{\text{RPA}}$  also increases, see Table~\ref{tab:PBEeigexc}.} The pole positions $\epsilon_{i}^\text{DFT}-\Omega_s^{\text{RPA}}$ are thus moving to the left on our energy scale. \reviewnew{ The pole at $-6.91~\text{eV}$, which was the culprit for the two solutions at the $G_0W_0@$PBE level, shifts now to $\varepsilon_{\text{HOMO-2}}^{\text{PBE0}}-\Omega_s^{\text{RPA@PBE0}} = - 9.70~\text{eV}$}. 
 A proper separation between satellite and QP peak is consequently also achieved at the $G_0W_0@$PBE0 level and we obtain also a unique QP solution for all five molecules. 
This confirms the observation from our ev$GW_0@$PBE study from Sec.~\ref{sec7}: our low-scaling $GW$ algorithm~\cite{wilhelm2021} is numerically precise if \reviewout{the correct physics is restored and} a well-defined QP peak is \reviewout{obtained} \reviewnew{present}.

\reviewnew{
$G_0W_0$ calculations starting from a hybrid functional with a larger fraction of exchange is common for molecules~\cite{Golze2019}. It was shown that 25\% of exact exchange, as used in PBE0, is optimal to minimize the deviation from CCSD(T) for frontier orbitals~\cite{caruso2016}. However, for our five molecules, 
the $G_0W_0$@PBE0 QP energies reported in Table~\ref{tab:pbe0} show worse agreement with CCSD(T) and experimental ionization potentials (Table~\ref{tab:spectral}) compared to ev$GW_0$@PBE. In addition, the $G_0W_0@$PBE0 approach is not extendable beyond molecular frontier orbitals: For deeper states, the amount of exact exchange needs to be progressively increased as excitation energies increase~\cite{Golze2020,Yao2022}. For solids, $G_0W_0$@PBE0 overestimates band gaps by more than 0.5 eV~\cite{Fuchs2007}. The $G_0W_0@$PBE0 behavior can be traced to the overestimation of excitation energies $\Omega_s$ of many-electron systems at the RPA@PBE0 level, see Table~\ref{tab:PBEeigexc}, Appendix~\ref{appb:Casida} and the discussion in  the literature, for example in Ref.~\cite{Golze2019}. Since $\Omega_s^\text{RPA@PBE0}$ energies are too large, the poles in the self-energy are shifted approximately to the same frequency range as for ev$GW_0$@PBE in case of the HOMO. However, it also leads to an underscreening in $W$, which affects the long-range interactions in solids more significantly than in finite molecular systems. 
%
%
%
%
%
%
}

\section{Conclusion}\label{sec8}
In summary, our study revisits low-scaling $GW$ calculations on molecules \ce{O3}, \ce{BeO}, \ce{MgO}, \ce{BN}, and \ce{CuCN} from the \gwoh benchmark set. 
 Previous studies~\cite{wilhelm2021,Azizi2024} encountered numerical challenges, particularly with the computation of the $G_0W_0$@PBE HOMO energy for these molecules from low-scaling $GW$. 
These challenges arise from \reviewout{the presence of} unphysical\reviewnew{, erroneous} multiple solutions of the quasiparticle equation at the $G_0W_0@$PBE level, which can be traced back to an insufficient separation between quasiparticle peak and satellite positions \reviewnew{caused by the too high PBE eigenvalues of occupied Kohn-Sham orbitals and the lack of eigenvalue-selfconsistency in~$G$.
This suggests that $G_0W_0@$PBE calculations of the HOMO of \ce{O3}, \ce{BeO}, \ce{MgO}, \ce{BN}, and \ce{CuCN} should be disregarded.}
Applying self-consistency in the Green's function (\evgwnospace) or using a hybrid functional as starting point for the $G_0W_0$ calculation  \reviewout{restores the correct separation between} \reviewnew{separates} QP peak and satellite\reviewout{. 
This leads} \reviewnew{leading} to a single HOMO QP peak and small satellite weights.
Low-scaling $GW$ computations yield numerically precise \reviewnew{\evgw and $G_0W_0$@PBE0} HOMO energies of \ce{O3}, \ce{BeO}, \ce{MgO}, \ce{BN}, and \ce{CuCN}, with a mean absolute deviation   in the order of only 10~meV from reference calculations\reviewout{with CD for $GW$  flavors which produce a physical result}. 
\reviewnew{We thus demonstrate that  low-scaling $GW$ is well-suited for computing frontier quasiparticle energies.}

\section*{Data availability}
Inputs and outputs of all calculations reported in this work  are available in a NOMAD repository~\cite{nomad_repo} and in a Github repository~\cite{github_repo}.
The low-scaling $GW$ algorithm~\cite{wilhelm2021} used in this work is available in the open-source package CP2K~\cite{cp2k,Kuehne2020}.

\begin{acknowledgments}
%
%
This work has been supported by the Innovation Study Exa4GW. 
Innovation Study Exa4GW has received funding through the Inno4scale project, which is funded by the European High-Performance Computing Joint Undertaking (JU) under Grant Agreement No 101118139. The JU receives support from the European Union's Horizon Europe Programme.
D.\,G.~and J.\,W.~acknowledge the Deutsche Forschungsgemeinschaft (DFG, German Research Foundation) for funding via the Emmy Noether Programme (Project No.~453275048 and 503985532, respectively). 
The authors gratefully acknowledge the computing time provided to them on the high performance computer Noctua 2 at the NHR Center PC2. These are funded by the Federal Ministry of Education and Research and the state governments participating on the basis of the resolutions of the GWK for the national high-performance computing at universities (www.nhr-verein.de/unsere-partner).

\end{acknowledgments}

\FloatBarrier
\appendix

\section*{Appendix}

\section{Spectral function and spectral weight}\label{sec:Z}
In this appendix, we derive Eq.~\eqref{qpweight} for the spectral weight~$Z_\varepsilon$ associated with a peak at energy~$\varepsilon$ of $A_n(\omega)$.
The spectral weight is the integral of that peak,
\begin{align} \label{Zeps}
    &Z_\varepsilon =  \int\limits_{\varepsilon-\Delta}^{\varepsilon+\Delta} A_n(\omega)\;\mathd \omega
    \\
 &
    =\int\limits_{\varepsilon-\Delta}^{\varepsilon+\Delta}\frac{1}{\pi}\,
    \frac{|
\gamma
    | }{\left(\omega -
  (\varepsilon + \text{Re} (\Sigma^\text{c}_n(\omega)-\Sigma^\text{c}_n(\varepsilon)
   )\right)^2  + 
  | \gamma|^2
   } \;\mathd \omega\,.
\label{Zepsgeneralintegral}
\end{align}
We use $\gamma\eqt \text{Im}\,\Sigma_n^\text{c}(\omega)\pt\eta $ and we integrate over the interval  $[\varepsilon\mt\Delta,\varepsilon\pt\Delta]$ with a parameter $\Delta\;{>}\;0$ to exclude other peaks from the integration. 
We have used Eq.~\eqref{An} and  the QP equation~\eqref{energy-it} to arrive at the denominator in Eq.~\eqref{Zepsgeneralintegral}. 
To  evaluate the integral~\eqref{Zepsgeneralintegral}, we 
  assume  that $\tmop{Im}\, \Sigma_n^\text{c} (\omega)$ is independent of frequency and we
use  Taylor expansion of $ \text{Re}\,\Sigma^\text{c}_n(\omega)$ around $\omega\eqt\epsilon$,
\begin{align}
 \text{Re}\Big[\Sigma^\text{c}_n(\omega)-\Sigma^\text{c}_n(\varepsilon) \Big]
 \approx \text{Re} \left. \frac{\partial \Sigma^\text{c}_n(\omega)}{\partial\omega }\right|_{ \displaystyle\omega=\varepsilon}\cdot (\omega-\epsilon)\,.
\end{align} 
Carrying out the integration~\eqref{Zepsgeneralintegral} gives the common approximate expression for the spectral weight, 
\begin{align*}
 Z_\varepsilon\approx \left[1-\text{Re}
  \left. \frac{\partial \Sigma^\text{c}_n(\omega)}{\partial\omega }\right|_{ \displaystyle\omega=\varepsilon}\right]^{-1}\,. 
\end{align*}

In practice, for Table~\ref{tab:spectral} and~\ref{tab:pbe0}, we determine the spectral weight~$Z$ of a peak in $A_\text{HOMO}(\omega)$ by fitting a Lorentz function 
\begin{align}
    \label{lorentz}
    L(\omega) = Z \cdot \frac{1}{\pi} \frac{\gamma}{(\omega - \epsilon_n^{G_0W_0})^2 + \gamma^2}
\end{align}
to  $A_\text{HOMO}(\omega)$.
We fix the position of the peak as the solution of the QP equation $\epsilon_n^{G_0W_0}$, while the peak width  $\gamma$ and the spectral weight $Z$ were fitted.

\reviewnew{Note that in non-selfconsistent $GW$ flavors like $G_0W_0$ or ev$GW_0$, the particle number is not conserved~\cite{Martin_Reining_Ceperley_2016}.
The violation is often in the order of 1\,\% or less~\cite{Martin_Reining_Ceperley_2016}.
For fully self-consistent $GW$, the particle number is conserved, i.e., $\int_{-\infty}^\infty \,{A_n^{GW}}(\omega) \,\text{d}\omega\eqt 1$.
}

\reviewnew{
\section{Excitation energies \(\Omega_s\) from Casida's equations}\label{appb:Casida}

In TDDFT, Casida's equations~\cite{Casida1995,Ullrich} provide a framework for calculating electronic excitation energies \(\Omega_s\). 
Casida’s formalism casts the TDDFT linear response problem into an eigenvalue problem,
\begin{align}
\left( \begin{array}{cc} \mathbf{A} & \mathbf{B} \\ -\mathbf{B} & -\mathbf{A} \end{array} \right) \left( \begin{array}{c} \mathbf{X}_s \\ \mathbf{Y}_s \end{array} \right) = \Omega_s^\text{TDDFT} \left( \begin{array}{c} \mathbf{X}_s \\ \mathbf{Y}_s \end{array} \right)\,,
\end{align}
where  \(\Omega_s^\text{TDDFT}\) are the excitation energies, i.e., the energy difference of an excited-state energy and the ground-state energy. 
$(\mathbf{X}_s,\mathbf{Y}_s)$ are the eigenvectors of the excitation.
The matrix elements of \(\mathbf{A}\) and \(\mathbf{B}\) are given by
\begin{align}
A_{ia,jb} &= (\epsilon_a^\text{DFT} - \epsilon_i^\text{DFT})\delta_{ij}\delta_{ab} + \langle ia | f_{\text{Hxc}} | jb \rangle \,,\label{eb2}
\\[0.3em]
B_{ia,jb} &= \langle ia | f_{\text{Hxc}} | bj \rangle\,,
\end{align}
where \(\epsilon_i^\text{DFT}\) and \(\epsilon_a^\text{DFT}\) are KS orbital energies,  \(\delta_{ij}\) and \(\delta_{ab}\) are Kronecker deltas and \(f_{\text{Hxc}} \eqt f_\text{H} \pt f_\text{xc}\) is the Hartree-exchange-correlation kernel, which includes both the Hartree and exchange-correlation (xc) kernel.
TDDFT in Casida's formulation is exact in case the exact xc kernel is used. 
In practice one relies on approximations, for example the adiabatic local-density approximation~\cite{Ullrich}, the  PBE functional~\cite{Perdew1996} or the PBE0 functional~\cite{Adamo1999}. 
This is important in $GW$ calculations as the self-energy~\eqref{e2} depends on excitation energies.
For small molecules, excitation energies~$\Omega_s^\text{TDDFT}$ computed from one of these approximations are often quite similar~\cite{Ullrich}; for MgO, we have 
\begin{align}
    \Omega_1^\text{PBE} = 0.67\,\text{eV}\approx\, 
\Omega_1^\text{PBE0} = 0.62\,\text{eV}\;, \label{eb4}
\end{align}
see also Table~\ref{tab:PBEeigexc}.
In contrast, the HOMO-LUMO gaps of PBE and PBE0 differ substantially; we have for the MgO molecule
\begin{align}
    \varepsilon^\text{PBE}_\text{LUMO}
    -
    \varepsilon^\text{PBE}_\text{HOMO}
&= 0.50\,\text{eV}    \,,
\\[0.5em]
    \varepsilon^\text{PBE0}_\text{LUMO}
    -
    \varepsilon^\text{PBE0}_\text{HOMO}
&=2.47\,\text{eV}  \,.
\end{align}

To better understand $\Omega_1^\text{PBE} \apt
\Omega_1^\text{PBE0} $ and  the big difference of the PBE and PBE0  HOMO-LUMO gap, we focus on large Coulomb interactions of densities with non-zero charge,
\begin{align}
\braket{nn|V|mm} = \int d\br\,d\br'\;   
\frac{|\psi_n(\br)|^2 \,|\psi_m(\br')|^2 }{|\br-\br'|}\,,
\end{align}
that are present in the exact-exchange part of PBE0.  
For an occupied orbital~$i$, the exact-exchange part of PBE0 contains the large interaction 
\begin{align}
 \Delta_i=  \beta\braket{ii|V|ii} \label{eb8}
\end{align}
where $\beta\eqt 0.25$ for PBE0. 
$\Delta_i$  reduces occupied KS orbital energies of small molecules  in PBE0 compared to PBE,
\begin{align}
   & \varepsilon_i^\text{PBE0} \approx  \varepsilon_i^\text{PBE}-\Delta_i\;.
\end{align}
This reduction of occupied orbital energies in PBE0 is a primary contribution to the increased  HOMO-LUMO gap in PBE0 compared to PBE.

Excitation energies from PBE and PBE0 are similar for small molecules, Eq.~\eqref{eb4}. 
We trace this finding back to the diagonal matrix elements of $A_{ia,ia}$, Eq.~\eqref{eb2}.
In PBE, we have $\langle ia | f_{\text{Hxc}}^\text{PBE} | ia \rangle \apt 0 $, such that $A_{ia,ia}\apt\varepsilon_a^\text{PBE}\mt\varepsilon_i^\text{PBE} $.
In PBE0, we have for small molecules~\cite{Ullrich} 
\begin{align}
\langle ia | f_{\text{Hxc}}^\text{PBE0} | ia \rangle     
\approx -\beta\braket{ii|V|aa}
\end{align}
where ${-}\braket{ii|V|aa}$ is the Coulomb interaction of an electron in a previously  empty orbital~$\psi_a(\br)$ with a hole in a previously occupied orbital~$\psi_i(\br)$.
In case orbitals are delocalized over the whole molecule, the Coulomb interaction $\braket{ii|V|aa}$ is similar to the Coulomb interaction of~$\psi_i(\br)$ with itself, $\braket{ii|V|aa}\apt \braket{ii|V|ii}$.
This allows us to use definition~\eqref{eb8} 
\begin{align}
\langle ia | f_{\text{Hxc}}^\text{PBE0} | ia \rangle     
\approx -\Delta_i\,,
\end{align}
such that we can estimate
\begin{align}
\begin{split}
    A_{ia,ia}^\text{PBE0}
    &\approx 
    \varepsilon_a^\text{PBE0} -  \varepsilon_i^\text{PBE0}
    -\Delta_i
\approx 
    \varepsilon_a^\text{PBE} -  \varepsilon_i^\text{PBE}
\approx 
     A_{ia,ia}^\text{PBE}\,.
     \end{split}\label{eb10}
\end{align}
Our estimate~\eqref{eb10} rationalizes $ 
    \Omega_1^\text{PBE} \apt
\Omega_1^\text{PBE0} $, Eq.~\eqref{eb4}.
Equation~\eqref{eb10} can be also understood by recognizing that the self-interaction in the HOMO level of KS-DFT and the absence of electron-hole Coulomb interaction in the xc kernel have similar magnitudes but opposite signs, leading to a mutual cancellation.

One often employs the RPA in $GW$, mainly for computational efficiency.
In RPA, the xc kernel \(f_\text{xc}\) is neglected in Casida’s equations,
\begin{align}&
\left( \begin{array}{cc} \mathbf{A}^\text{RPA} & \mathbf{B}^\text{RPA} \\ -\mathbf{B}^\text{RPA} & -\mathbf{A}^\text{RPA} \end{array} \right) \left( \begin{array}{c} \mathbf{X}^\text{RPA}_s \\ \mathbf{Y}^\text{RPA}_s \end{array} \right) = \Omega_s^\text{RPA} \left( \begin{array}{c} \mathbf{X}^\text{RPA}_s \\ \mathbf{Y}^\text{RPA}_s \end{array} \right),
\\
&A_{ia,jb}^\text{RPA} = (\epsilon_a^\text{DFT} - \epsilon_i^\text{DFT})\delta_{ij}\delta_{ab} + \langle ia | f_{\text{H}} | jb \rangle\,,\label{eb12}
\\[0.3em]
&B_{ia,jb}^\text{RPA} = \langle ia | f_{\text{H}} | bj \rangle\,.
\end{align}
When using the PBE functional, RPA leaves the diagonal elements of $\mathbf{A} $ almost unchanged because $\langle ia | f_{\text{xc}}^\text{PBE} | ia \rangle \apt 0 $, 
\begin{align}
 A_{ia,ia}^\text{RPA@PBE} \approx  
     \varepsilon_a^\text{PBE} -  \varepsilon_i^\text{PBE}
     \approx  A_{ia,ia}^\text{ PBE} 
        \approx  A_{ia,ia}^\text{ PBE0}\, .
\end{align}
Thus, PBE excitation energies change only little under RPA; for the MgO molecule, we have
\begin{align}
    \Omega_1^\text{RPA@PBE}\hspace{0.3em} &= 0.82\,\text{eV} \approx \hspace{0.2em} \Omega_1^\text{PBE} \;\;= 0.67\,\text{eV} \;.
\end{align}

When applying the RPA with the PBE0 functional,
the diagonal elements of $\mathbf{A}$   increase by $\apt\Delta_i$:
\begin{align}
\begin{split}
    A_{ia,ia}^\text{RPA@PBE0}
    &\overset{\eqref{eb12},\eqref{eb10} }{\approx} 
    \varepsilon_a^\text{PBE0} -  \varepsilon_i^\text{PBE0}
\approx 
     A_{ia,ia}^\text{PBE0}+\Delta_i\,.
     \end{split}
\end{align}
This increase translates to spuriously increased excitation energies in RPA@PBE0; again, for MgO, we have 
\begin{align}
\Omega_1^\text{RPA@PBE0} &= 2.83\,\text{eV}\gg
 \Omega_1^\text{PBE0} = 0.62\,\text{eV} \;,\label{eb17}
\end{align}
see also Table~\ref{tab:PBEeigexc}.
RPA excitation energies are used in  the $GW$ self-energy, see Eq.~\eqref{e2}.
It has already long been recognized that starting a $G_0W_0$ calculation from PBE0 features too large RPA@PBE0 excitation energies~\cite{Golze2019}, see Eq.~\eqref{eb17} as an example.
It is often recommended to use the PBE functional as starting point for $GW$ calculations because of $\Omega_s^\text{RPA@PBE}\apt \Omega_s^\text{PBE}$. 
To cure the too high PBE eigenvalues of occupied KS orbitals in $G$, Eq.~\eqref{e2}, it is often recommended to use ev$GW_0$@PBE~\cite{Golze2019}.
}

\section{Self-energy and spectral function of all molecules \ce{O3}, \ce{BeO}, \ce{MgO}, \ce{BN}, \ce{CuCN}}
\begin{figure*}\vspace{-1em}
    \centering    \includegraphics[width=\textwidth]{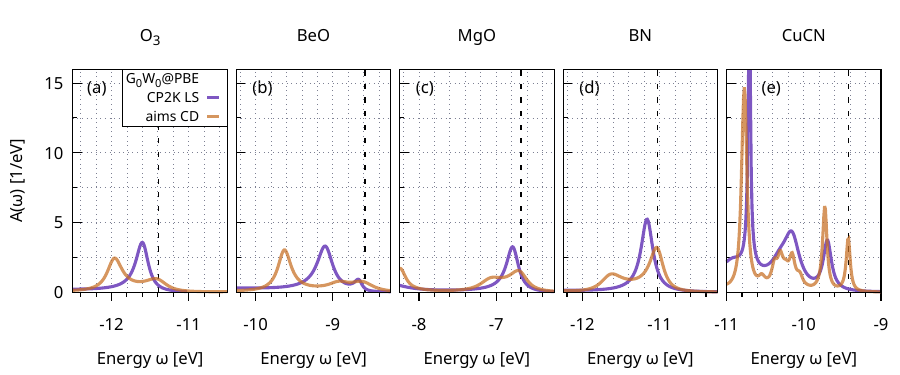}
    \caption{(a)-(e) $G_0W_0$@PBE spectral function  for \ce{O3}, \ce{BeO}, \ce{MgO}, \ce{BN}, \ce{CuCN} computed from low-scaling $GW$ in CP2K~\cite{wilhelm2021} (CP2K LS) and from CD in FHI-aims~\cite{Golze2018} (aims CD).
    The dashed lines indicate the $G_0W_0$@PBE HOMO energy computed from FHI-aims and 128-paramater Padé continuation taken from the original \gwoh work~\cite{gw100}. }
    \label{fa1}
\end{figure*}

\begin{figure*}\vspace{2em}
    \centering
    \includegraphics[width=1.0\textwidth]{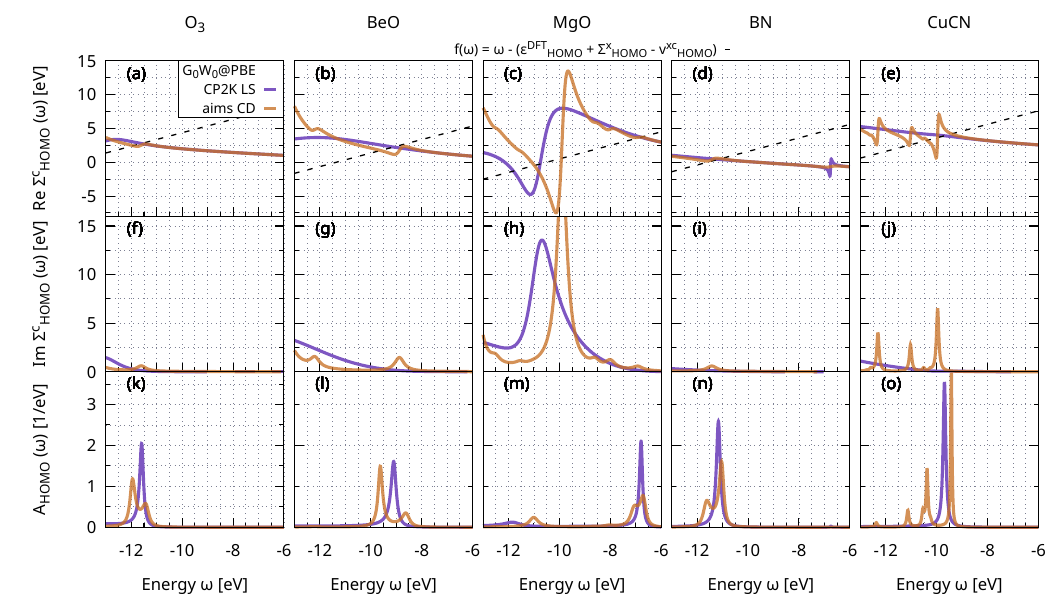}
    \caption{ (a)-(e) Re($\Sigma^\text{c}_\text{HOMO}$), (f)-(j) Im($\Sigma^\text{c}_\text{HOMO}$), (k)-(o) $A_\text{HOMO}$ [Eq.~\eqref{An}]
    for \ce{O3}, \ce{BeO}, \ce{MgO}, \ce{BN} and CuCN computed from $G_0W_0$@PBE. 
    }
    \label{fa2}
\end{figure*}

\begin{figure*}
    \centering
    \includegraphics[width=1.0\textwidth]{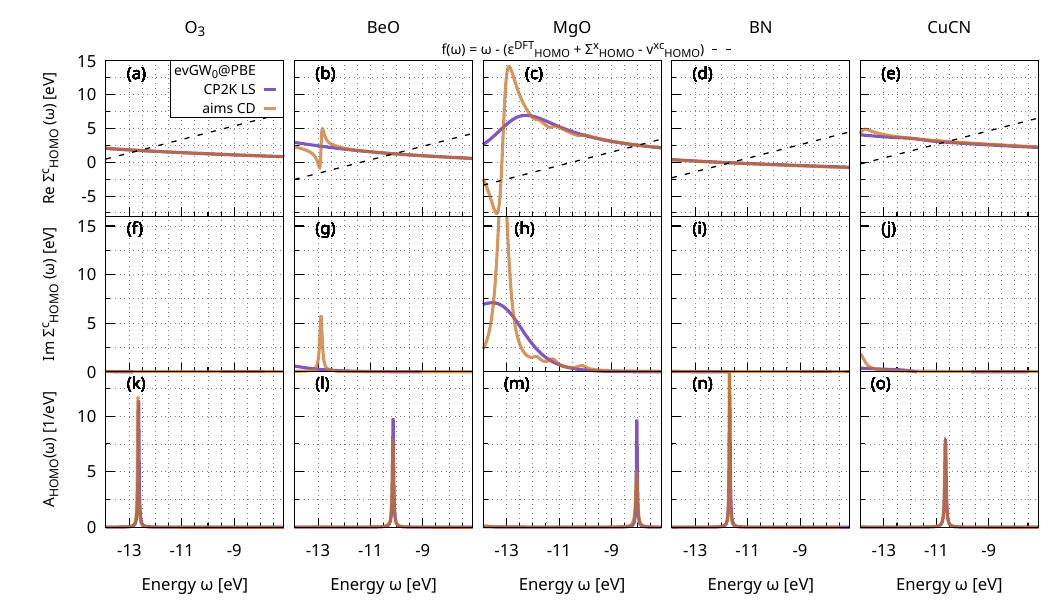}
    \caption{ (a)-(e) Re($\Sigma^\text{c}_\text{HOMO}$), (f)-(j) Im($\Sigma^\text{c}_\text{HOMO}$), (k)-(o) $A_\text{HOMO}$ [Eq.~\eqref{An}]
    for \ce{O3}, \ce{BeO}, \ce{MgO}, \ce{BN} and CuCN computed from \evgwnospace@PBE. 
}
    \label{fa3}
\end{figure*}

\begin{figure*}
    \centering
    \includegraphics[width=1.0\textwidth]{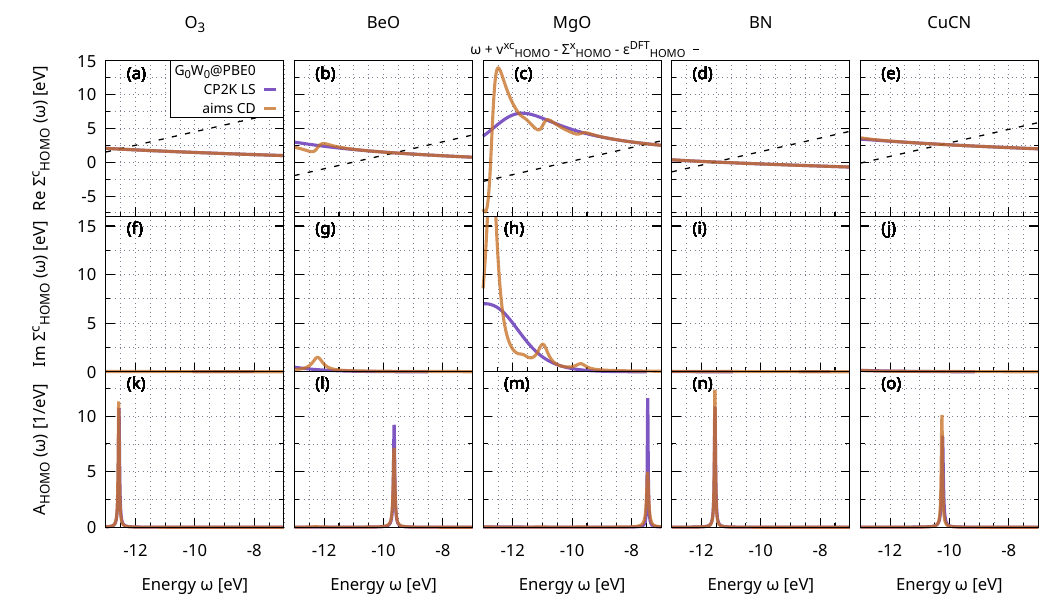}
    \caption{ (a)-(e) Re($\Sigma^\text{c}_\text{HOMO}$), (f)-(j) Im($\Sigma^\text{c}_\text{HOMO}$), (k)-(o) $A_\text{HOMO}$ [Eq.~\eqref{An}] for
    \ce{O3}, \ce{BeO}, \ce{MgO}, \ce{BN} and CuCN computed from $G_0W_0$@PBE0. 
    }
        \label{fa4}
        \vspace{-3em}
\end{figure*}

We provide spectral function~$A(\omega)$ of the previously challenging five molecules \ce{O3}, \ce{BeO}, \ce{MgO}, \ce{BN}, \ce{CuCN} computed from  $G_0W_0$@PBE (Fig.~\ref{fa1}).
We also show
Re$(\Sigma^\text{c}_\text{HOMO})$, Im$(\Sigma^\text{c}_\text{HOMO}$) and $A_\text{HOMO}$ [Eq.~\eqref{An}] for \ce{O3}, \ce{BeO}, \ce{MgO}, \ce{BN} and CuCN computed from $G_0W_0$@PBE,  ev$GW_0$@PBE and $G_0W_0$@PBE0 (Fig.~\ref{fa2}, \ref{fa3} and~\ref{fa4}).


\newpage
\FloatBarrier

\bibliography{main}

\end{document}